\documentclass[aps,showpacs]{revtex4}
\usepackage{graphicx}
\usepackage{float}
\usepackage[T1]{fontenc}
\usepackage[latin1]{inputenc}
\usepackage{amssymb}
\include{epsf}
\epsfverbosetrue

\newcommand{\xx}{\noindent}

\newcommand{\bea}{\begin{eqnarray}}
\newcommand{\eea}{\end{eqnarray}}

\newcommand{\colzero}{\left (\begin{array}{c}1\\0\end{array}\right )}

\newcommand{\colone}{\left (\begin{array}{c}0\\1\end{array}\right )}

\makeatother
\begin{document}

\title{The Dynamics of Entanglement in the Adiabatic Search and Deutsch Algorithms}

\author{K. Choy$^\dagger$, G. Passante$^{*}$, D. Ahrensmeier$^\dagger$, M.E. Carrington$^\dagger$, T Fugleberg$^\dagger$, R. Kobes$^*$ and G. Kunstatter$^*$}

\affiliation{$^\dagger$ Department of Physics, Brandon University,
Brandon, Manitoba, R7A 6A9 Canada\\
and Winnipeg Institute for Theoretical Physics,
Winnipeg, Manitoba, Canada}

\affiliation{$^*$ Physics Department, University of Winnipeg, 
Winnipeg, Manitoba, R3B 2E9 Canada\\
and Winnipeg Institute for Theoretical Physics,
Winnipeg, Manitoba, Canada}

\begin{abstract}
The goal of this paper is to study the effect of entanglement on the running time of a quantum computation. 
Adiabatic quantum computation is suited to this kind of study, since it allows us to explicitly calculate the time evolution of the entanglement throughout the calculation. On the other hand however, the adiabatic formalism makes it impossible to study the roles of entanglement and fidelity separately, which means that results have to be interpreted carefully. 
We study two algorithms: the search algorithm and the Deutsch-Jozsa algorithm. We find some evidence that entanglement can be considered a resource in quantum computation.
\end{abstract}


\date{\today}

\maketitle

\section{Introduction}
In the past few years, it has become increasingly clear that entanglement is  an important resource in quantum information science.  It is crucial for quantum teleportation \cite{teleportation}, and has been proven to be necessary for an exponential speed-up of quantum computation over classical computation \cite{Jozsa0201143,Vidal0301063,Azuma0102}. However, there is some uncertainty over the precise role that entanglement plays in general. It seems not to be
necessary for improvements less dramatic than exponential. Lloyd \cite{Lloyd9903057} has shown that quadratic speed-up can be achieved with no entanglement whatsoever, using single particle states, albeit at the very significant cost of an exponential increase in what he called ``overhead'', eg. the number of particle detectors required. A recent implementation of Grover's search algorithm using only classical waves \cite{Bhattacharya} suggests that in this case interference is the relevant factor. Finally, it has been shown by Biham {\it et al} \cite{Biham0306182} using specific examples that quantum computers  without entanglement can still provide improvements that cannot, even in principle, be classically achievable . 

One of the complications in analyzing the role of entanglement, as Lloyd's analysis, for example, indicates, is that there are many other factors which affect the speed of a quantum computation.  Examples are the Hilbert-space dimension (e.g. in \cite{Caves0304083}), 
and even the measurement process itself \cite{Castagnoli0005069}. In the case of adiabatic quantum computation, it was shown \cite{Das0204044} that a temporary, large input of energy into the system
could also  increase the computation speed. This is similar to the 
results of Lloyd \cite{Lloyd9903057} cited above, with recoverable energy playing the role 
of ``overhead''. Finally, we mention the connection highlighted by Wei  {\it et al} \cite{Wei0504113} between the minimal
running time of an adiabatic computation and the  fidelity (distance) between the initial and final state in the adiabatic algorithm.

In light of all these results, it is of great interest to analyse as fully as possible the precise role that entanglement plays in achieving speed-up of quantum algorithms.
In the following we will focus on the dynamical role of entanglement in adiabatic quantum computing. Adiabatic quantum computation provides an alternative model for quantum computation which considers a system evolving under a time-dependent Hamiltonian instead of through a sequence of gates. It has recently been shown to be equivalent to standard quantum computation with respect to computational power \cite{Aharonov0405}, and may be more promising when it comes to robustness against decoherence or other types of errors (e.g. \cite{Childs0108}). For our purposes it is an ideal theoretical framework in which to study the role of entanglement and fidelity, since one can track the actual evolution of these quantities as a continuous function of time.

Whereas some work has been done studying the entanglement dynamics for ``naturally occurring'' many-body systems, e.g. a Heisenberg chain \cite{DeChiara0512586}, two spins in a magnetic field \cite{Novaes0510162}, or two qubits in a thermal bath \cite{Lucamarini0402073}, the behaviour of entanglement in systems while a quantum algorithm is running has only been investigated for a few special cases: Orus and Latorre \cite{Orus0311017} calculated the time dependence of entanglement for two adiabatic quantum algorithms (NP-exact cover problem and Grover's search problem). In the former, which runs in polynomial time, they found indications that the entanglement grows linearly with the number $n$ of qubits. In the latter case, which achieves only a quadratic speed-up, they showed analytically that the maximum entanglement is bounded in the limit of large $n$. Finally, Roland and Cerf \cite{rc0302138} have studied the fidelity as a function of time in the locally adiabatic search algorithm. They showed that for the optimized algorithm, which achieves a $\sqrt{N}$ speed-up, the fidelity is linear in the optimized time variable, in keeping with the fact that the standard quantum search algorithm changes the fidelity in equal increments. 

In addition to shedding light on the specific role of entanglement in the efficiency of quantum algorithms, a systematic investigation of the time evolution of entanglement during adiabatic quantum computation may also provide other benefits. Firstly, to achieve a better understanding of the engineering of quantum information processes, it is necessary to understand the dynamics of the entangled states which are used for the quantum computation.
On a more fundamental level, new insight from quantum information theory may lead to a better understanding of interacting many-body systems, and phenomena like quantum phase transitions and critical behaviour (see, e.g., \cite{Meyer0308082}). 

 
We must, at this stage, emphasize an inherent limitation associated with the study of the time evolution of entanglement during an adiabatic quantum computation. We generally start from a set of initial states that have zero initial entanglement but different initial fidelity relative to the desired final state. We expect that the initial fidelity (which measures the separation of the initial state from the final state) and the amount of entangement that is produced during the evolution, will both affect the running time of the algorithm.   Because the Hamiltonian, and therefore the algorithm, depends explicitly on the choice of initial state, it is not possible to study the role of each variable separately. Consequently, it is difficult to determine to what extent each is influencing the other. 

We study two different algorithms, adiabatic quantum search and the adiabatic Deutsch-Jozsa algorithm \cite{Das0111032}. We also discuss a variation of the Deutsch-Jozsa algorithm proposed in \cite{Wei0512008} which we will refer to as the constant-time-Deutsch-Jozsa algorithm. To study the possible dependence of our results on the definition of entanglement, we also introduce a new function to describe the physical property of entanglement and compare the results with those for the entropy of entanglement. Similarly, we compare the results for the fidelity with those for a different function characterizing the distance between states. We also calculate the time evolution of these quantities as a function of an optimized time variable, which corresponds to applying the adiabatic approximation in a maximally efficient way  \cite{rc0302138}.

For the search algorithm, using the unoptimized time variable, we find that the region where the magnitude of the rate of change of the fidelity (as defined in (\ref{fid})) is greatest corresponds to the region of greatest entropy production. Using the optimized time variable, the $n=2$ initial states that produce more entanglement have longer running times. For $n=3$ this correlation only holds when we use the new function introduced in this paper to measure entanglement. 
For Deutsch's algorithm, using the optimized time variable, we find that states that generate entropy more quickly have a shorter running time. The constant-time-Deutsch-Jozsa algorithm behaves like the search algorithm for $n=2$:  an increased production of entropy is correlated with a longer running time. An earlier form of some of these results was described in \cite{Ahrensmeier0512066}. \\

Our paper is organized as follows: In section (\ref{adiabatic-section})
we review adiabatic quantum computation and the algorithms that we will study. 
In section (\ref{entanglement-section}) we review the entropy of entanglement and the fidelity,
and introduce two related, but new, definitions of functions to describe the same  physical 
properties. The results are presented in section (\ref{results}) and we present our conclusions in 
section (\ref{conc}). Appendix (\ref{AppendixA}) contains some notational 
definitions.  In Appendix (\ref{AppendixB}) we derive expressions
for the reduced density matrices.  In Appendices (\ref{AppendixC}),(\ref{AppendixD})   
and (\ref{AppendixE})
we present some details of the calculations related to a new definition of entanglement. 
\section{Adiabatic Quantum Computation}
\label{adiabatic-section}
Standard quantum computation uses quantum circuits, in which a sequence of unitary transformations acts on a set of qubits. The time evolution takes place in discrete steps. The building blocks of an adiabatic quantum computer are qubits, too, but the time evolution of the system is described by a time-dependent Hamiltonian, and is therefore much closer to the way a physicist describes interacting (many-body) systems.

Calculations for a system with a time dependent Hamiltonian are much more involved than for a time-independent one, but fortunately several approximation methods exist for different limiting cases. The adiabatic approximation applies to systems in which the Hamiltonian varies sufficiently slowly in time. Roughly speaking, this approximation says that if a system is in an instantaneous eigenstate and its Hamiltonian evolves slowly enough in time, the system will remain in this instantaneous eigenstate. The eigenstate and its energy eigenvalue evolve continuously, but there is no transition to a different state. (We have restricted our considerations here to the most simple case, assuming that the spectrum of $H(t)$ is entirely discrete and nondegenerate. In the case of a degenerate spectrum, the adiabatic approximation generalizes to the statement that each degenerate eigenspace, instead of the individual eigenvectors, evolves independently \cite{Sarandy0405059}). This situation is intuitively plausible, since we know that for a time-independent Hamiltonian, the system would stay in a given eigenstate for all times. The approximation is made quantitative  by the adiabatic theorem, which we briefly sketch below, following the excellent presentation in \cite{Sarandy0405059}. 
\subsection{Adiabatic Theorem}
Consider a system obeying a time-dependent Schr\"{o}dinger equation
\begin{equation}\label{schrgl}
 i \frac{d|\Psi(t)\rangle}{dt}=H(t)|\Psi(t)\rangle,
\end{equation}
where here and in the following we set $\hbar =1$.
The time-dependent Hamiltonian has instantaneous eigenstates $|E_n(t)\rangle$ which satisfy
\begin{equation}
\label{EIGEN}
 H(t)|E_n(t)\rangle = E_n(t)|E_n(t)\rangle.
\end{equation}
Note that both the energies and the state vectors are functions of time. To simplify the notation we will suppress the `$t$' and denote the ground state and its energy by $\{|E_-\rangle\,,\;E_-\}$, the first excited state and its energy by $\{|E_+\rangle\,,\;E_+\}$, and higher states and energies by $\{|E_n\rangle\,,\;E_n\}$.  
Using these states as a basis, we can express a general solution of (\ref{schrgl}) as
\begin{equation}
 |\Psi(t)\rangle=\sum_{n}a_n(t)e^{i\alpha_n(t)}|E_n(t)\rangle,
\end{equation}
where
\begin{equation}
\alpha_n(t)=-\int_0^t E_n(t')dt'
\end{equation}
is the dynamical phase. After some algebraic manipulations, we obtain the following expression for the time evolution of the coefficients:
\begin{equation}
\label{coefficients}
 \dot{a}_k=-a_k\langle E_k|\dot{E_k}\rangle -\sum_{n\neq k}a_n
 \frac{\langle E_k|\dot{H}|E_n\rangle}{E_n(t)-E_k(t)}e^{i(\alpha_n - \alpha_k)}.
\end{equation}

The condition for adiabatic evolution is that the coefficients $a_k(t)$ evolve independently from each other, i.e. that their dynamical equations do not couple. This is ensured if the following condition is satisfied:
\begin{equation}
\label{cond1}
\max_{0\leq t\leq T}\left|
\frac{\langle E_k|\dot{H}|E_n\rangle}{E_n-E_k}\right|\ll\min_{0\leq t\leq T}|E_n-E_k|
\end{equation}
where $T$ is the total evolution time. Physically, this means that if the time evolution of the Hamiltonian (in units of the energy gap between the levels under consideration) is slow compared to this energy gap, for all pairs of energy levels, the condition for adiabaticity is fulfilled. 

If we consider specifically the adiabatic evolution of an initial state which is an eigenvector, 
\begin{equation}
\label{initialc}
|\Psi(0)\rangle =|E_m(0)\rangle,
\end{equation}
then $a_m(0)=1$ and $a_n(0)=0\quad \forall n\neq m$, and the evolution equation (\ref{coefficients}) reduces to
\begin{equation}
\dot{a}_m=-a_m\langle E_m|\dot{E}_m\rangle.
\end{equation}
The adiabatic evolution of the state is 
\begin{equation}\label{adiabatic}
 |\Psi(t)\rangle = e^{i(\alpha_m(t)+\gamma_m(t))}|E_m(t)\rangle,
\end{equation}
where the geometrical phase (or Berry phase) $\gamma_m(t)$ is given by
\begin{equation}
\gamma_m(t)=i\int_0^t dt'\langle E_m(t')|\dot{E_m}(t')\rangle.
\end{equation}
This phase depends on the path in parameter space that is traversed by the time-dependent parameters in $H$. Eqs. (\ref{initialc}) and (\ref{adiabatic}) state the content of the adiabatic theorem: An instantaneous eigenstate $|E_m(0)\rangle$ evolves continuously to the corresponding eigenstate $|E_m(t)\rangle$ at a later time, without any transition to other energy levels.

When the time dependence of the Hamiltonian is not small enough to fulfill the condition for adiabaticity, transitions to other eigenstates become allowed. Equivalently, the dynamical equations (\ref{coefficients}) will couple distinct eigenvectors, and the eigenvectors will not evolve completely independently. The time evolution of such a system could be obtained by solving the system of coupled differential equations (\ref{coefficients}). 

\subsection{The Hamiltonian for adiabatic quantum computation}
\label{ham-section}
Adiabatic quantum computation takes the system from an initial state $|\psi_0\rangle$ to a final state $|\psi_1\rangle$, which are eigenstates of the Hamiltonians $H_0$ and $H_1$, respectively. The initial state is assumed to be easy to construct, for practical purposes, and the final state encodes the solution to the computational problem, which is unknown. The time dependent Hamiltonian is constructed so that it will drive the system from the initial to the final state,  moving slowly enough to ensure 
adiabatic evolution, so that it will end up in the desired state with large probability. This Hamiltonian is usually constructed as a linear combination of the initial and final Hamiltonians,

\bea
\label{ham0}
H(t) = f(t)H_0+g(t)H_1.
\eea
The functions $f(t)$ and $g(t)$ are arbitrary  functions of time, usually chosen to be monotonic,  which are subject to the boundary conditions
\bea
&& f(0)=1\,;~~g(0) = 0 \\
&& f(T)=0\,;~~g(T) = 1 
\eea
where $T$ represents the total computation time. For example, the simplest choice of these functions is 
\bea
\label{simplefg}
f(t) = 1-s(t)\,;~~g(t) = s(t)\,;~~s(t) = t/T
\eea
where we have introduced the dimensionless time variable $s(t)$.
The initial and final Hamiltonians are constructed from the initial and final states as
\bea
\label{ham2}
&& H_0 = I - |\psi_0\rangle\langle\psi_0| \\
&& H_1 = I - |\psi_1\rangle\langle\psi_1| \nonumber
\eea
so that the ground state of $H_0$ is the state $|\psi_0\rangle$ and the ground state of $H_1$ is the state $|\psi_1\rangle$. 
\subsection{Running time of the algorithm}
\label{running-time}
We define a small dimensionless parameter corresponding to (\ref{cond1}):
\bea
\label{epsilon}
\epsilon = \max_{0\leq t\leq T}
\left|\frac{\langle E_-|\dot{H}|E_+\rangle}{(E_+-E_-)^2}\right|.
\eea
 For $\epsilon\ll 1$, the adiabatic approximation is satisfied. 
Rewriting (\ref{epsilon}) in terms of the dimensionless variable $s=t/T$ and assuming that the bound is saturated we obtain:
\bea
\label{tRUN}
T = \max_{0\leq s\leq 1}
\frac{\left|\langle E_-|\frac{dH}{ds}|E_+\rangle\right|}{\epsilon(E_+-E_-)^2}.
  \eea
 We obtain the running time by choosing a particular value for $\epsilon$, which represents the accuracy of the calculation.   Physically, for times greater than $T$ the system can be assumed to be in the final state $|\psi_1\rangle$ (see Eq. (\ref{ham2})), up to errors of order $\epsilon^2$. 
 
 The minimal running time can be obtained by optimizing the adiabatic approximation \cite{rc0302138}. Effectively, we apply the adiabatic condition (\ref{epsilon}) to each infinitesimal time step $dt$ throughout the evolution, instead of globally to the entire time interval $T$. We replace the function $s(t) = t/T$ with a non-linear function whose slope is large in regions where the energy gap is large, and small when the energy gap is small. The function $s(t)$ is obtained by solving the differential equation:
\bea
\label{Trun}
dt =  ds \left|
\frac{\epsilon(E_+-E_-)^2}{\langle E_-|\frac{dH}{ds}|E_+\rangle}\right| 
\eea
We choose $\epsilon = 0.01$  and obtain the optimized running time as:
\bea
T_{op}= (0.01) \cdot \int_0^1 ds \left|
\frac{(E_+-E_-)^2}{\langle E_-|\frac{dH}{ds}|E_+\rangle}\right|
\eea

\subsection{Adiabatic Quantum Search Algorithm}
The goal of the search algorithm is to find a marked item, the ``needle,'' in a large unsorted database, the ``haystack,'' in as few steps as possible. Classically, $O(N)$ steps are needed for a database of size $N$, since all the objects have to be examined, one after another. For a quantum search, the items are represented by the computational basis states, one of them being the marked state $|m\rangle$. The unsorted database is represented by an equally weighted superposition of these states:
\bea
\label{isQS} 
|\psi_0\rangle = \frac{1}{\sqrt{N}}\sum_{i=0}^{N-1}|i\rangle
\eea
which is motivated by the idea that if a measurement were made on the initial state, the probability of selecting any item would be the same. 
The state $|\psi_0\rangle$ is taken as the initial state for the calculation, and the final state is $|\psi_1\rangle = |m\rangle$, the marked state that we are searching for. Without loss of generality, we choose $|m\rangle =|0\rangle$. In addition to using the conventional ``haystack'' initial state, we perform the numerical calculation using several other choices. Specifically, we vary the initial state  in order to study the correlation between running time, entanglement and fidelity.


For the adiabatic quantum search, the Hamiltonian is constructed from the initial and final state as in Eqs.(\ref{ham0}) and (\ref{ham2}) (see \cite{Das0204044} or \cite{Roland02}). 
Its matrix elements are
\begin{eqnarray}
H_{11} &= & f(1-\frac{1}{N})\\
H_{ii} &= & f(1-\frac{1}{N})+g\qquad\mbox{for}\, i\ne 1\nonumber\\
H_{ij} & = & -\frac{f}{N} \qquad\mbox{for} \, i\ne j\nonumber
\end{eqnarray}
The lowest two eigenvalues are
\bea
E_\pm(t) = \frac{1}{2}\left((f+g)\pm \sqrt{(f-g)^2+\frac{4}{N}f\,g}\right),
\eea
and the only  higher energy eigenvalue $E_n=f+g$ is $(N-2)$-fold degenerate \cite{Das0204044}. To calculate the running time, we only need to determine explicitly the lowest eigenvector. Within the adiabatic approximation, this lowest eigenvector gives the instantaneous state of the system and is used in the determination of the entanglement and fidelity as functions of time.
\subsection{Deutsch-Jozsa Algorithm}
The Deutsch-Jozsa problem \cite{DeutschJozsa} is to determine whether a Boolean function 
\begin{equation}
F:\{0,1\}^n\rightarrow \{0,1\}
\end{equation}
is constant (i.e. all outputs are identical) or balanced (i.e. it has an equal number of zeroes and ones as output). Classically this requires $\sim O(2^n/2)$ evaluations of the function. Quantum mechanically one can start with an equal superposition of all values of the input variable and determine the solution using a single application of the oracle operator $U_f$ that is the quantum analogue of the classical function evaluation. Although this is a somewhat artificial example, it provides an excellent illustration of how quantum superpositions allow maximal use of the information content of the full Hilbert space. What is less clear is whether entanglement plays any role in this case.

 For the simplest case of $n=1$, the function would be constant if $F(0)=F(1)$, and balanced if  $F(0)\neq F(1)$. The four possible outcomes in this case are
\bea
&& F(0) = F(1) = 0 ~({\rm constant}) \\
&& F(0) = F(1) = 1 ~({\rm constant}) \nonumber\\
&& F(0) = 0;~F(1) = 1 ~({\rm balanced}) \nonumber\\
&& F(0) = 1;~F(1) = 0 ~({\rm balanced}) \nonumber
\eea

Classically one has to determine both $F(0)$ and $F(1)$ to infer the nature of the function, since the knowledge of one does not shed any light on the value of the other. Using the quantum version, Deutsch's algorithm \cite{Deutsch85}, the nature of the function can be determined by making one measurement.
 
An adiabatic realization of the Deutsch-Jozsa algorithm has been introduced in \cite{Das0111032}. The initial state is the same as in the search algorithm, Eq. (\ref{isQS}). Unlike the case of the search algorithm, there is no natural physical motivation for this initial state in Deutsch's algorithm, but it has been shown in \cite{Das0111032} that this choice is optimal in the sense that the time evolution is independent of the values of $\alpha$ and $\beta$ (defined below), and thus independent of the result of the algorithm. For $n$ qubits, the running time for this algorithm scales as $\sqrt{N}=2^{n/2}$.  Once again, we perform the numerical calculation using several other choices for the initial state, in order to study the correlation between running time, and entanglement and fidelity. 

The final state is given by
\begin{equation}
|\psi_1\rangle=\alpha |0\rangle+\frac{\beta}{\sqrt{N-1}}\sum_{k=1}^{N-1}|k\rangle,
\end{equation}
where 
\begin{eqnarray}
 \alpha & = & \frac{1}{N}\left | \sum_{x \in \{0,1\}^n} (-1)^{F(x)} \right |\\
 \beta & = & 1 - \alpha.\nonumber
\end{eqnarray}
 If $F$ is constant, $\alpha =1$ and $\beta =0$, and if $F$ is balanced, $\alpha =0$ and $\beta =1$. This means that if the measurement of the final state after the required running time yields $|0\rangle$, the function is constant, otherwise it is balanced. 
 
The Hamiltonian is given by:
\bea
H_{11} & = & \frac{f(N-1)}{N}+g(1-\alpha)\\
H_{ii} & = & \frac{f(N-1)}{N}+g(1-\frac{\beta}{N-1})\,\mbox{for}\,i\ne 1\nonumber\\
H_{1i} & = & H_{i1} = -\frac{f}{N}\,\mbox{for}\,i\ne 1\nonumber\\
H_{ij} &= &-\frac{f}{N}-\frac{g \beta}{(N-1)}
 \,\mbox{for}\,i\ne j,i\ne 1,j\ne 1\nonumber
\eea

The two lowest energy eigenvalues are
\begin{equation}
E_{\pm}(s)=\frac{1}{2}\left ( 
(f+g)\pm\sqrt{(f-(\alpha -\beta)g)^2+\frac{4}{N}(\alpha - \beta)fg} \right ),
\end{equation}
and the higher eigenvalue $E_n=f+g$ is $(N-2)$-fold degenerate \cite{Das0111032}. Note that the minimum energy gap between the lowest two states is:
\bea
g_{min}\equiv |E_+-E_-|_{min} \sim O(1/\sqrt{N})
\eea
which goes to zero for large $N$, and explains heuristically the scaling of the running time.
\subsection{Constant-Time-Deutsch-Jozsa Algorithm}
Since the standard Deutsch-Jozsa algorithm requires only one call of the oracle operator, it is natural to ask whether one can do better than a $\sqrt{N}$ improvement over the classical case. This question has been answered in the affirmative in \cite{Wei0512008} where the authors were able to construct an algorithm that solved the problem in constant time, independent of $N$.  In order to achieve this, they start with the same initial state, but choose the final state to have the form
\bea
|\psi_1\rangle = \frac{\alpha}{\sqrt{N/2}}\sum^{N/2-1}_{i=0}|2i\rangle +\frac{\beta}{\sqrt{N/2}}\sum^{N/2-1}_{i=0}|2i+1\rangle
\eea
This version of Deutsch's algorithm is optimized in the sense that the running time can be shown to be independent of $N$. The Hamiltonian is given by:
\bea
&& (H_0)_{mn} = \delta_{mn}-\frac{1}{N}\sum^{N-1}_{\{j,k\}=0}\delta_{mj}\delta_{nk}\\
&& (H_1)_{mn} = \delta_{mn} -\frac{2\alpha}{N}\sum^{N/2-1}_{\{j,k\}=0}\delta_{m(2j)}\delta_{n(2k)} -\frac{2\beta}{N}\sum^{N/2-1}_{\{j,k\}=0}\delta_{m(2j+1)}\delta_{n(2k+1)}\nonumber
\eea
In this case, a straightfoward calculation reveals the two lowest energy eigenvalues to be:
\bea
E_{\pm} = {1\over 2} \pm {1\over 2} \sqrt{1-2s(1-s)}
\eea
This yields a minimum energy gap that is independent of $N$ which at a heuristic level accounts for the fact that the running time is independent of $N$.
\section{Quantifying the possible resources: Entanglement and Fidelity}
\label{entanglement-section}


\subsection{Entropy of entanglement}
There is no generally accepted measure for the entanglement of an arbitrary state. 
The exception is the case of a bipartite system in a pure state. 

For a bipartite system, composed of subsystems A and B, the Hilbert space is given by a tensor product $\cal{H}_A\otimes\cal{H}_B$. The state of the composite system is a product state if it can be written as a product of pure states of subsystems, 
\begin{equation}
|\Psi_{AB}\rangle =|\Psi_A\rangle \otimes |\Psi_B\rangle .
\end{equation}
If the state of the system cannot be written as a product state, it is entangled.  An example of an entangled state for the simplest bipartite system, a system of two qubits, is the singlet Bell state
\bea
|\Psi\rangle_{{\rm Bell}}^0 = \frac{1}{\sqrt{2}}\Big(|0\rangle_A \otimes |1\rangle_B-|1\rangle_A\otimes |0\rangle_B\Big).
\eea

The von Neumann entropy as a measure of entanglement makes use of the fact that for an entangled pure state, the state of the composite system is known completely, but the states of the subsystems are not (they are not in pure states). It
measures the amount of information about one subsystem  that can be obtained by making a measurement on the other subsystem. It is calculated from the reduced density matrix of either of the subsystems:
\begin{eqnarray}
\label{ent}
 S_{vN}(\rho_{red}) &&= -{\rm Tr}\,(\rho_{red} \log_2 \rho_{red})\\
              & & =-\sum_n \mu_n\log_2 \mu_n,\nonumber
\end{eqnarray}
where the $\mu_n$ are the eigenvalues of the reduced density matrix. If the composite state is a product state, it has zero entanglement and the reduced density matrix will have eigenvalues zero and one. From Eq. (\ref{ent}) the von Neumann entropy will be zero. If the composite system is not in a product state, the von Neumann entropy will be non-zero. Its maximum value for a bipartite system is 1.

If the system is in a pure state, but is multipartite, then it is not clear how  it should be partitioned. In other words, the von Neumann entropy can depend on which qubits are traced out. It is unclear what physical significance to attach to an entanglement measure that depends on the specific partitioning of the system.
For a multipartite system, or for mixed states, several other, inequivalent measures of entanglement have been suggested. Although there are general conditions that need to be satisfied by any viable measure (see, e.g., \cite{Vedral97} for a discussion), no generally accepted unique definition exists as yet.
Several authors have considered many-particle systems, e.g. spin-systems, as bipartite by simply
splitting them in the middle \cite{Orus0311017}. Another possibility is to consider one qubit in the
``mean field'' generated by the other qubits, as it is done, for example, in \cite{Latorre0304098}.  

In this paper we study systems with $n=2$ and $n=3$ qubits. We give explicit expressions for the eigenvalues of the reduced density matrix for $n=2$ qubits and $n=3$ qubits in the Appendix B. 
\subsection{Measuring  entanglement geometrically}
\label{Gt}
In this section we suggest a different function to measure the entanglement of a state. This definition essentially measures the distance between the composite state under consideration and the product state that is closest, in a space parametrized by the coefficients of this product state. 
We note that this is a physically motivated approach, different from the mathematical or axiomatic \cite{Vedral97}, or the operational approach (as suggested in \cite{entmeasures}). One advantage of the definition that we propose in this section is that its generalization for the case of more than two subsystems is straightforward: the problem of how to partition the system does not occur. For simplicity of presentation we discuss only real cofficients. 
 
We begin by discussing the Hilbert-Schmidt measure of entanglement $D_{HS}$, which is defined as the minimum (over the set $S$ of product states) of the Hilbert-Schmidt distance between the density matrix $\rho_c$ of the state under consideration and the density matrix $\rho_p$  of a product state \cite{Bertlmann0508043}:
\begin{equation}
 D_{HS}(\rho_c)=\min_{\rho\in S} d_{HS}(\rho_p,\rho_c),
\end{equation}
where the Hilbert-Schmidt distance is defined as
\begin{equation}
d_{HS}(\rho_p,\rho_c)=|| \rho_p-\rho_c ||
\;;~~
|| \rho ||=\sqrt{Tr \rho^{\dagger}\rho}.
\end{equation}
To simplify the notation, we discuss below the case $n=2$. 
A general product state of two qubits can be written:
\begin{eqnarray}
\label{2qubit}
 |p\rangle & = & (a_1|0\rangle +a_2|1\rangle)\otimes (b_1|0\rangle +b_2|1\rangle)\\
 & = & a_1b_1|00\rangle +a_1b_2|01\rangle+a_2b_1|10\rangle+a_2b_2|11\rangle\nonumber\\
 & = & (a_1b_1,a_1b_2,a_2b_1,a_2b_2)\nonumber
\end{eqnarray}
(Note that throughout this paper we write the components of vectors as rows of the form $(v_1,\;v_2,\;v_3,\;\cdots)$, instead of as columns, to save space). 
The function to be minimized is then (the square root of)
\begin{equation}
\label{HS}
d_{HS}^2(a_1,a_2,b_1,b_2) = 2-2(a_1b_1c_0+a_1b_2c_1+a_2b_1c_2+a_2b_2c_3)^2.
\end{equation}

Using this definition as a starting point, we define a new function as follows.
The difference between the state under consideration, $|c\rangle$, and a product state is given by:
\begin{equation}
 |d\rangle =|c\rangle -|p\rangle =(c_0-a_1b_1,c_1-a_1b_2,c_2-a_2b_1,c_3-a_2b_2)
\end{equation}
Our entanglement measure is the minimum of the distance $\langle d|d\rangle$ for all product states. The function to be minimized is:
\begin{eqnarray}
\label{distance}
D(a_1,a_2,b_1,b_2) & = & \langle d | d \rangle \\
     & = &  c_0^2+c_1^2+c_2^2+c_3^2 + (a_1^2+a_2^2)(b_1^2+b_2^2)\nonumber\\
     & &  -2(a_1b_1c_0+a_1b_2c_1+a_2b_1c_2+a_2b_2c_3),\nonumber
\end{eqnarray}
where the state under consideration is taken to be normalized, which gives $c_0^2+c_1^2+c_2^2+c_3^2=1$. The minimum of this function with respect to the parameters of the product state is a measure of how much the state $|c\rangle$ differs from a product state, and thus a measure of entanglement.
One advantage of our definition $D$ in comparison with the Hilbert-Schmidt distance is that it is faster to calculate numerically, basically because of the absence of crossterms (compare Eqs. (\ref{HS}) and (\ref{distance})). The advantage over the von Neumann entropy is that there is no ambiguity about how to partition a multi-partite system. 

The next step is to minimize the function (\ref{distance}) with respect to the parameters of the product state. We discuss this procedure below for the case $n=2$. In Appendix C we consider $n=3$. We will consider two different procedures. The first method uses normalized product states only. This approach is motivated by the fact that physical states are normalized. 
The normalization condition for the product state is
\begin{equation}
(a_1b_1)^2 +(a_1b_2)^2+(a_2b_1)^2+(a_2b_2)^2 = N_a \cdot N_b =1
\end{equation}
with $N_a=a_1^2+a_2^2,N_b=b_1^2+b_2^2$. We do not assume that the one-qubit states are individually normalized.
For the case of $n=2$ we have obtained analytic expressions for the values of the parameters $\{a_1,a_2,b_1,b_2\}$ that correspond to the product state closest to any arbitrary state $|c\rangle$. These expressions are given in Appendix D. We have also obtained numerical solutions for $n=3$. These results are shown in section (\ref{results}).

We also study the minimum of (\ref{distance}) using unnormalized product states. In this case it is easy to show that, in order find a non-trivial minimum of the distance function, one must satisfy a condition on the normalization constants $N_a,N_b$. We obtain this condition by looking at the four equations
\bea
\frac{\partial D}{\partial a_1} = \frac{\partial D}{\partial a_2}=\frac{\partial D}{\partial b_1}=\frac{\partial D}{\partial b_2}=0
\eea
and rearranging terms. The result is that, in order to find a non-trivial solution, we must satisfy the condition:
\begin{equation}
\label{consis}
 (N_a N_b)^2 -N_aN_b +|c_0c_3-c_1c_2|^2 =0
\end{equation}
We note that this equation has the same form as the equation from which we obtain the eigenvalues of the reduced density matrix (see Appendix B.1):  the equation ${\rm det}(\rho^{red}_2-\mu)=0$ has the form
\bea
 \mu^2 -\mu +|c_0c_3-c_1c_2|^2 =0 \,.
\eea
This result suggests that for $n=2$ there is a connection between entanglement as defined through the von Neumann entropy, and entanglement defined as the minimum distance to an unnormalized product state. To understand this connection, it is useful to consider a geometric interpretation (see Appendix D for details). The space of all normalized 4-dimensional states $|c\rangle$ is a  3-sphere. The space of normalized product states is a subset on the surface of this 3-sphere. 
Unnormalized product states lie along radial lines that intersect the 3-sphere.
The closest product state to an arbitrary normalized state will lie on such a radial line in the interior of the 3-sphere {\it ie.} it will be an unnormalized product state. The only case where the closest product state will be normalized is if the original arbitrary state is itself a product state.  This follows directly from Eq. (\ref{consis}), which implies that $N_aN_b<1$ unless $|c_0c_3-c_1c_2|^2=0$.
Since the quantity $N_a N_b$ is the length of the 4-dimensional product state vector and it satisfies the same equation as the eigenvalues of the reduced density matrix, Eq.(\ref{consis}) provides a geometric interpretation of one of these eigenvalues.  Therefore, one of the terms of the von Neumann entropy for an arbitrary normalized state is determined by the length
of the closest non-normalized product state. The equations that are analogous to (\ref{consis}) for $n=3$ are found in Appendix C.

We obtain numerical solutions for $D_{min}$ as a function of time, using both normalized and unnormalized product states. In all cases, the results are almost identical. We show both graphs for the case of the $n=2$ search algorithm only. 
\subsection{Fidelity}
The fidelity measures the closeness of two states, in the sense defined below. The general definition of the fidelity of two states is written in terms of their density matrices $\rho$ and $\sigma$ as \cite{Nielsen}
\begin{equation}
\tilde F(\rho,\sigma)=tr\sqrt{\sqrt{\rho}\sigma\sqrt{\rho}}.
\end{equation}
For pure states $\rho = |i\rangle\langle i|$ and $\sigma = |j\rangle\langle j|$, the fidelity is just the overlap
\begin{equation}\label{fidelity}
\tilde F(|i\rangle,|j\rangle)=|\langle i|j\rangle |
\end{equation}
of the two states. In our case, we are interested in measuring the closeness of the instantaneous eigenstate $|c(t)\rangle$ and the final state $|c(T)\rangle$. We look at the function
\bea
\label{fid}
F(t) = 1-\tilde F(|c(t)\rangle,|c(T)\rangle) = 1-\Big|\sum_{i=0}^{N-1}c^*_i(t)c_i(T)\Big|
\eea
which approaches zero as $t\rightarrow T$. Introducing a slight abuse of terminology, we henceforth refer to $F$ as the 
fidelity. Using this convention, the fidelity approaches zero as the
desired state is reached (i.e. in the limit $t\to T$).
\subsection{Another distance measure}
There are several other natural definitions for the distance between the instantaneous ground state $|c(t)\rangle$ and the final state $|c(T)\rangle$. We propose  the following function:
\begin{equation}
\label{G}
G(t)=\sum_{i=0}^{N-1} \left ||c_i(t)|-|c_i(T)| \right |
\end{equation}
\section{Calculations and Results}
\label{results}

We work with a set of initial states that have zero initial entanglement. Our goal is to study the relationship between  the running time of the algorithm and the amount of entanglement that is produced during the evolution of the algorithm. As described in section (\ref{ham-section}), the initial state determines the Hamiltonian $H(s)$ for each algorithm as a function of the dimensionless time variable $s=t/T$. The eigenvectors and eigenvalues can be obtained numerically at each point in time. From these expressions we can extract the coefficients $\{c_0,~c_1,\cdots c_{N-1}\}$ that correspond to the components of the eigenvectors in the computational basis (see Eqns (\ref{2-gen}) and (\ref{3-gen})). Using these coefficients we calculate the von Neumann entropy from (\ref{ent}), (\ref{2-mu}), (\ref{3-mu}); the geometric entanglement as the minimum of (\ref{HS}); the fidelity from (\ref{fid}); and the geometric separation from (\ref{G}). These quantities are obtained numerically as functions of the time $s$ throughout the evolution of the system from the initial state $|\psi_0\rangle$ at $s=0$ to the final state $|\psi_1\rangle$ at $s=1$.

We also present some results in terms of the optimized time $t$. As explained in section (\ref{running-time}), the optimized time is obtained by applying the adiabatic condition (\ref{epsilon}) to each infinitesimal time step $dt$ throughout the evolution, instead of globally to the entire time interval $T$. The function $s(t) = t/T$ is replaced with a non-linear function that is calculated numerically. 

We note that the search algorithm and Deutsch's algorithm with $(\alpha=1,~\beta=0)$ are identical since we can choose (without loss of generality) the marked state of the search algorithm to be $|m\rangle=|0\rangle$. As a consequence, we only have to investigate the case $(\alpha=0,~\beta =1)$ for Deutsch's algorithm. We also note that for the search algorithm and the constant time Deutsch algorithm, the system evolves to a final state which has zero entanglement (since the final state is a product state). On the other hand, Deutsch's algorithm with $\beta=1$ evolves to a final state with non-zero entanglement.

The initial states that we use, for $n=2$ and $n=3$, are given below:
\bea
&&n=2 \\
&&|{\rm red}\rangle = {\rm Normalized}\{1,3/2,1,3/2\} \nonumber\\
&&|{\rm yellow}\rangle = {\rm Normalized}\{1,1,4/3,4/3\} \nonumber\\
&&|{\rm green}\rangle = {\rm Normalized}\{1,1,1,1\} \nonumber\\
&&|{\rm blue}\rangle = {\rm Normalized}\{1,1,2/3,2/3\} \nonumber\\
&&|{\rm cyan}\rangle = {\rm Normalized}\{2,2,1,1\} \nonumber\\
&&|{\rm magenta}\rangle = {\rm Normalized}\{3,1,3,1\} \nonumber\\[4mm]
&&n=3\\
&&|{\rm red}\rangle = {\rm Normalized}\{1,1,3,3,1,1,3,3\} \nonumber\\
&&|{\rm yellow}\rangle = {\rm Normalized}\{3,3,3,3,1,1,1,1\} \nonumber\\
&&|{\rm green}\rangle = {\rm Normalized}\{1,1,1,1,1,1,1,1\} \nonumber\\
&&|{\rm blue}\rangle = {\rm Normalized}\{1,1,2,2,1,1,2,2\} \nonumber\\
&&|{\rm cyan}\rangle = {\rm Normalized}\{2,2,2,2,3,3,3,3\} \nonumber\\
&&|{\rm magenta}\rangle = {\rm Normalized}\{19,19,1,1,19,19,1,1\} \nonumber
\eea

\xx The general features of our results are as follows:\\

\xx {\sc [A] Consistency checks on the calculation:}

\xx (1) The optimized running time is less than the unoptimized running time.

\xx (2) The running time is larger where the initial fidelity is larger.

\xx (3) The same behaviour is seen for the geometric separation.\\

\xx {\sc [B] Search Algorithm:}

\xx (1) Using the unoptimized time, the region where the magnitude of the rate of change of the fidelity is greatest corresponds to the region of greatest entropy.

\xx (2) Using the optimized time, the
speed of the algorithm is adjusted so that it moves faster at either end of the time interval and more slowly in the middle, where the gap is smallest. This effectively flattens out the curve that gives the rate of change of the fidelity. 

\xx (3) Using optimized time, states with a larger maximum entanglement tend towards longer running times. We remind the reader that the interpretation of this result is unclear, because of the inherent structure of the adiabatic calculation: different initial states have different initial fidelity, and thus it is not clear how to separate the effects of fidelity and entanglement on the running time. In other words, it seems likely that the longer running time is actually caused by the larger initial fidelity, and not by the larger maximum entanglement. This conclusion is supported by point (B.1) which indicates that increased entanglement causes the fidelity to move more quickly towards its final value. \\

\xx {\sc [C] Deutsch Algorithm:}

\xx (1) Using the optimized time variable we find that states that generate entropy more quickly have a shorter running time. \\

\

\xx {\sc [D] Constant Time Deutsch Algorithm:}

\xx (1) Using the unoptimized time, the region where the magnitude of the rate of change of the fidelity is greatest corresponds to the region of greatest entropy. 

\xx (2) Using optimized time, states with a larger maximum entanglement tend towards longer running times. As discussed for point (B-3) above, it seems likely that the longer running time is actually caused by the larger initial fidelity, and not by the larger maximum entanglement. This conclusion is supported by point (D.1).\\

\xx We give below the numerical results and graphs that illustrate these points.

\newpage

\subsection{Search Algorithm for two qubits}

The running times are (point (A-1)):
\bea
T_{run}\Big|_{\rm unoptimized} &&:~~ \{T_{\rm red} = 598,~ T_{\rm yellow} = 503,~ T_{\rm green} =346,~ T_{\rm blue} = 234,~ T_{\rm cyan} = 194,~ T_{\rm magenta} = 165\}\nonumber\\
T_{run}\Big|_{\rm optimized} &&:~~ \{T_{\rm red} = 234,~ T_{\rm yellow} = 213,~ T_{\rm green} =173,~ T_{\rm blue} = 137,~ T_{\rm cyan} = 122,~ T_{\rm magenta} = 110\}\nonumber
\eea
\par\begin{figure}[H]
\begin{center}
\includegraphics[width=8cm]{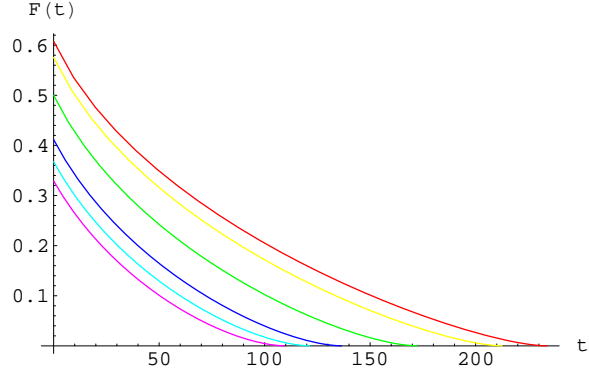}
\end{center}
\caption{Largest initial fidelity corresponds to the largest running time (point (A-2)).}
\end{figure}
\par\begin{figure}[H]
\begin{center}
\includegraphics[width=8cm]{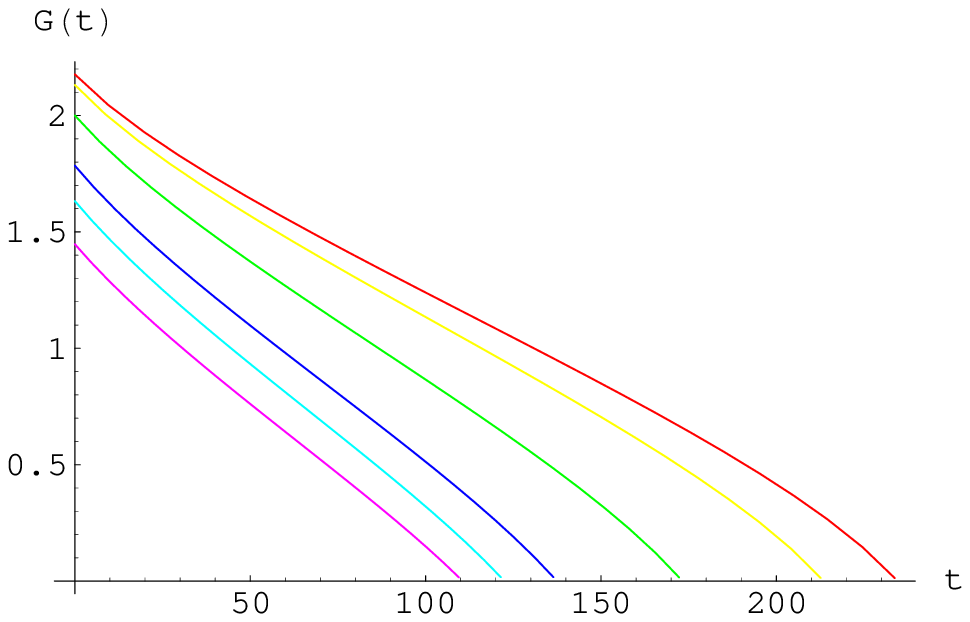}
\end{center}
\caption{Largest initial geometric separation corresponds to the largest running time (point (A-3)).}
\end{figure}
\par\begin{figure}[H]
\begin{center}
\includegraphics[width=8cm]{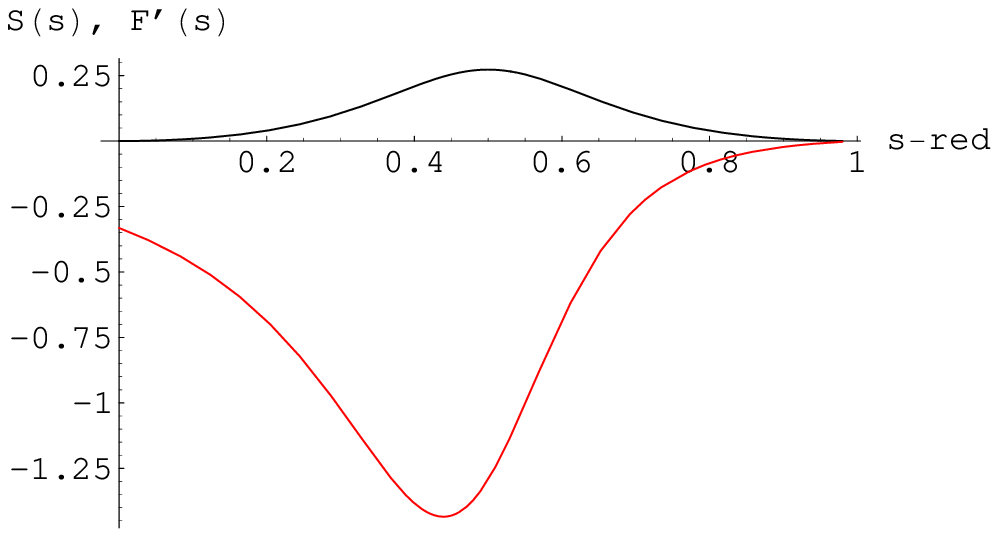}
\end{center}
\caption{The region of greatest magnitude of the rate of change of the fidelity  corresponds to the region of greatest entanglement (point (B-1)). This graph is for the red initial state.}
\end{figure}
\par\begin{figure}[H]
\begin{center}
\includegraphics[width=8cm]{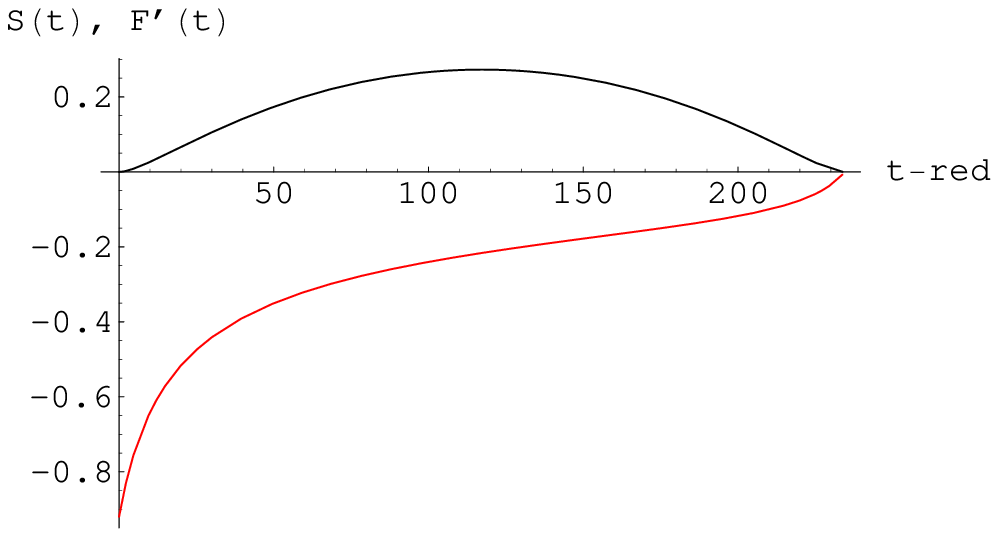}
\end{center}
\caption{The rate of change of the fidelity  as a function of the optimized time is flattened (point (B-2)). This graph is for the red initial state.}
\end{figure}
\par\begin{figure}[H]
\begin{center}
\includegraphics[width=8cm]{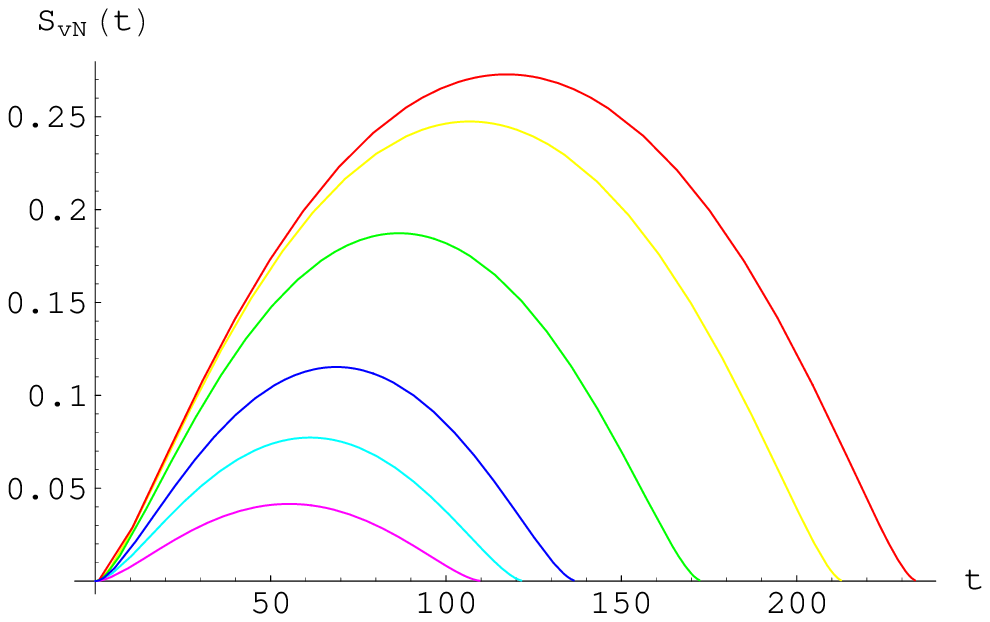}
\end{center}
\caption{Entropy versus optimized time (point (B-3)).}
\end{figure}
\par\begin{figure}[H]
\begin{center}
\includegraphics[width=8cm]{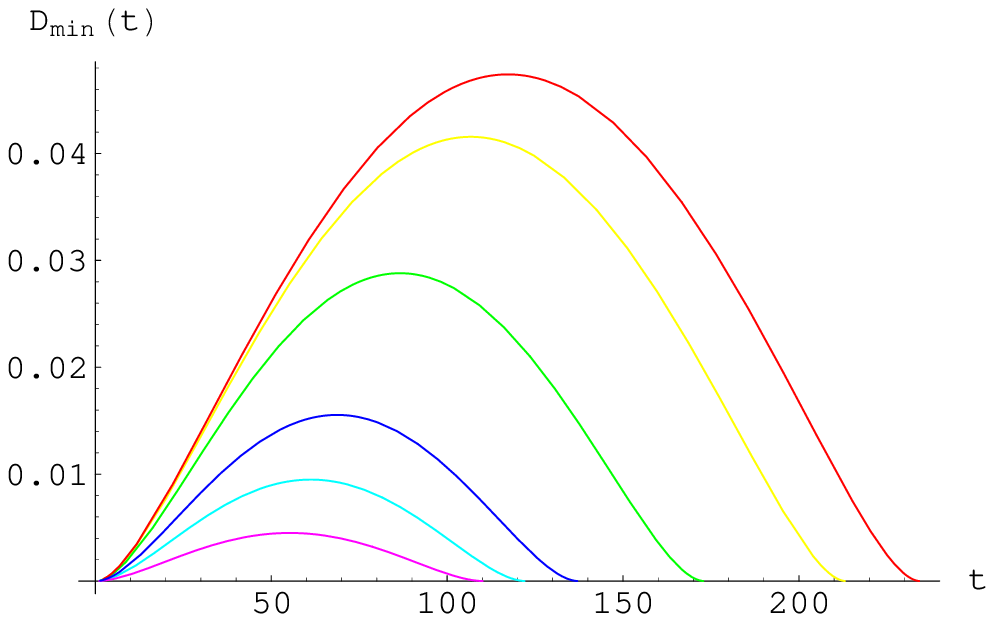}
\end{center}
\caption{Normalized geometric separation versus optimized time (point (B-3)).}
\end{figure}
\par\begin{figure}[H]
\begin{center}
\includegraphics[width=8cm]{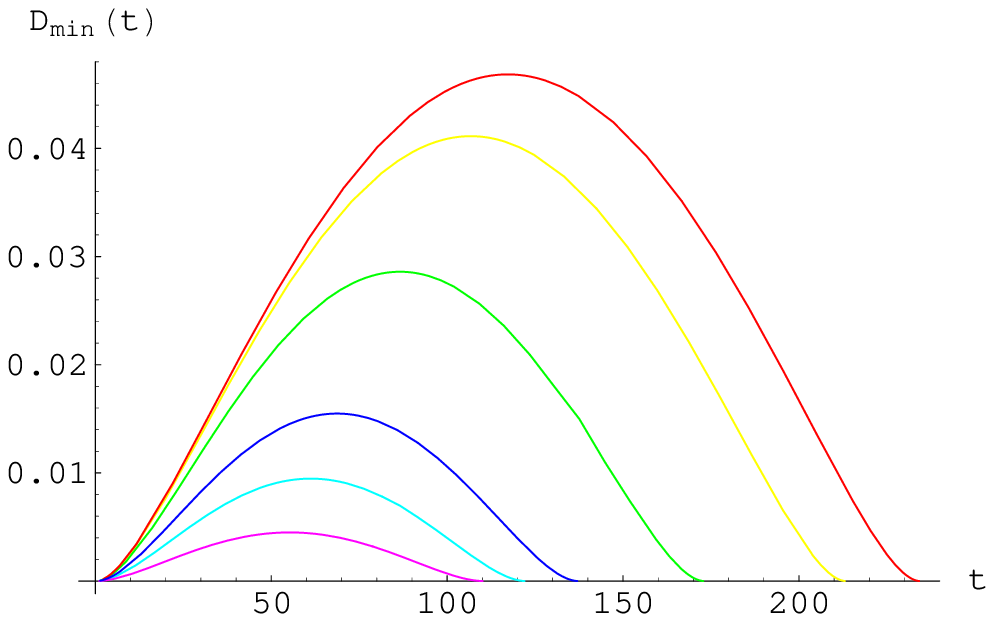}
\end{center}
\caption{Unnormalized geometric separation versus optimized time (point (B-3)).}
\end{figure}
\newpage

\subsection{Search Algorithm for three qubits}

The running times are (point (A-1)):
\bea
T_{run}\Big|_{\rm unoptimized} &&:~~ \{T_{\rm red} = 3950,~ T_{\rm blue} = 1949,~ T_{\rm cyan} =1249,~ T_{\rm green} = 748,~ T_{\rm yellow} = 391,~ T_{\rm magenta} = 348\}\nonumber\\
T_{run}\Big|_{\rm optimized} &&:~~ \{T_{\rm red} = 624,~ T_{\rm blue} = 436,~ T_{\rm cyan} =346,~ T_{\rm green} = 264,~ T_{\rm yellow} = 185,~ T_{\rm magenta} = 173\}\nonumber
\eea
\par\begin{figure}[H]
\begin{center}
\includegraphics[width=8cm]{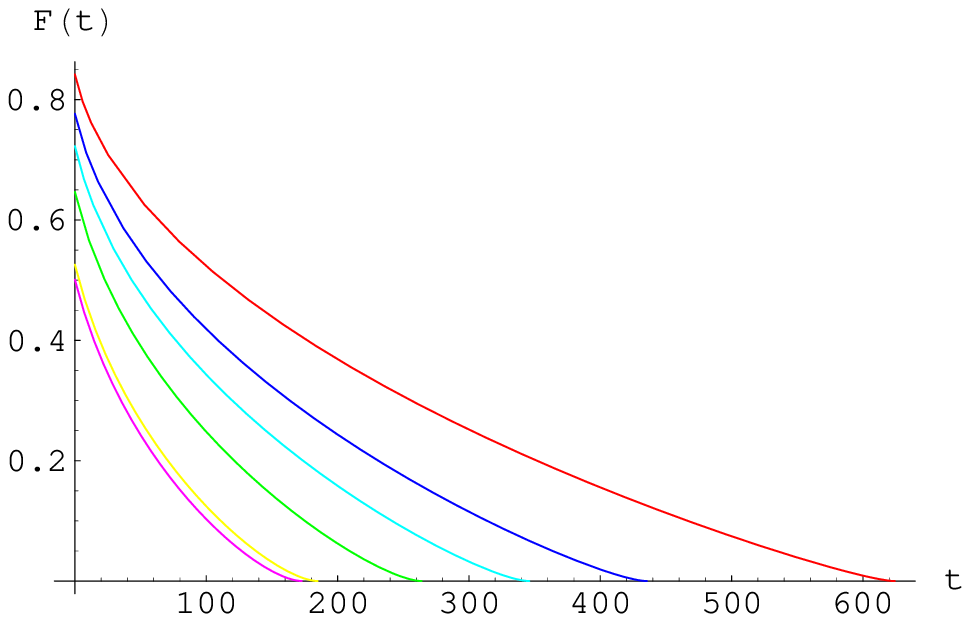}
\end{center}
\caption{Largest initial fidelity corresponds to the largest running time (point (A-2)).}
\end{figure}
\par\begin{figure}[H]
\begin{center}
\includegraphics[width=8cm]{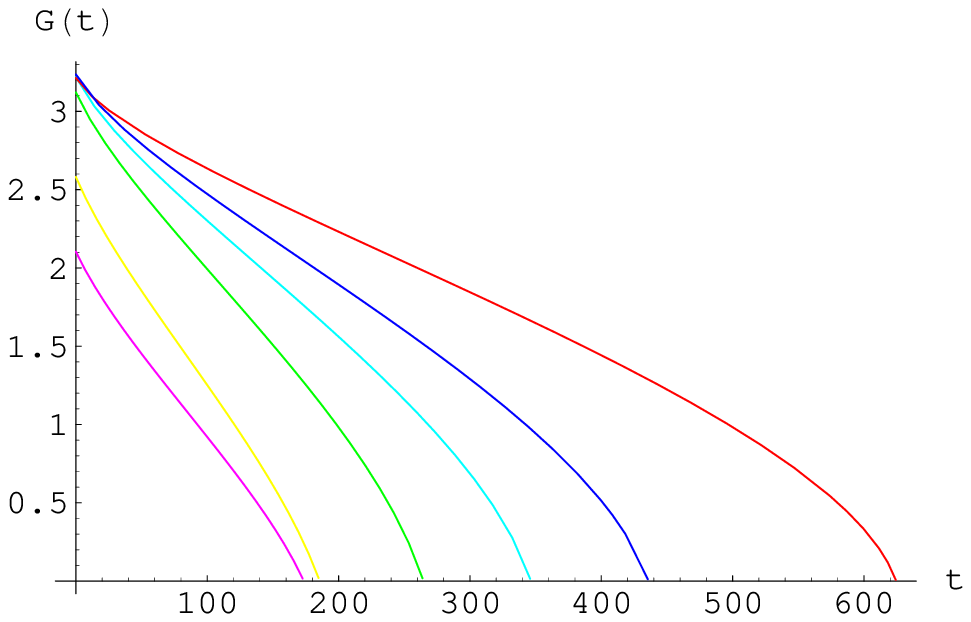}
\end{center}
\caption{Largest initial geometric separation corresponds to the largest running time (point (A-3)).}
\end{figure}
\par\begin{figure}[H]
\begin{center}
\includegraphics[width=8cm]{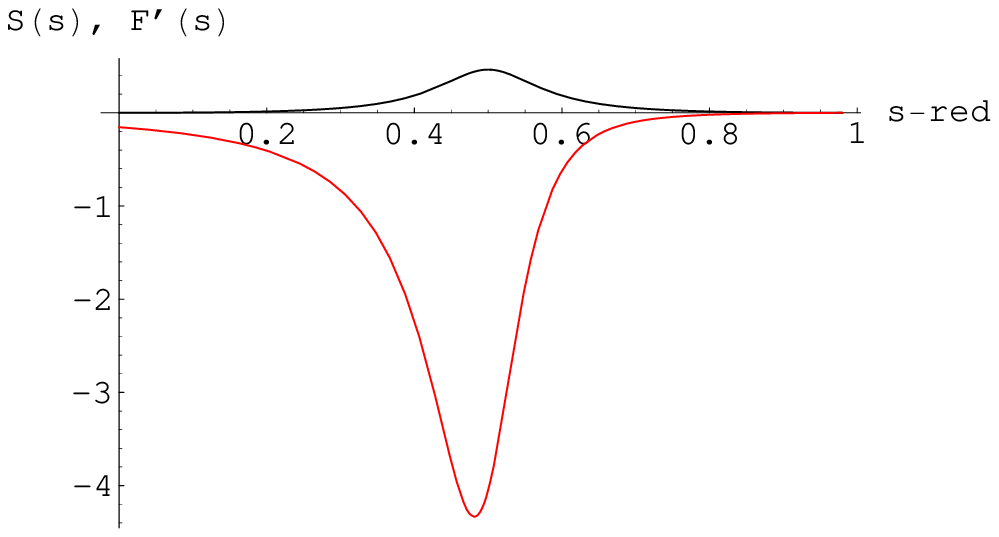}
\end{center}
\caption{The region of greatest magnitude of the rate of change of the fidelity  corresponds to the region of greatest entanglement (point (B-1)). This graph is for the red initial state.}
\end{figure}
\par\begin{figure}[H]
\begin{center}
\includegraphics[width=8cm]{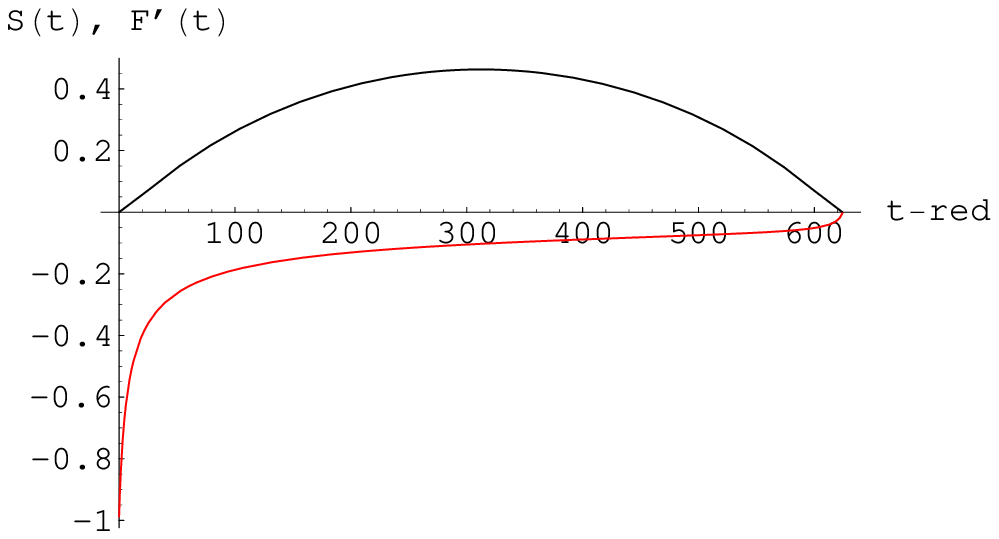}
\end{center}
\caption{The rate of change of the fidelity  as a function of the optimized time is flattened (point (B-2)). This graph is for the red initial state.}
\end{figure}
\par\begin{figure}[H]
\begin{center}
\includegraphics[width=8cm]{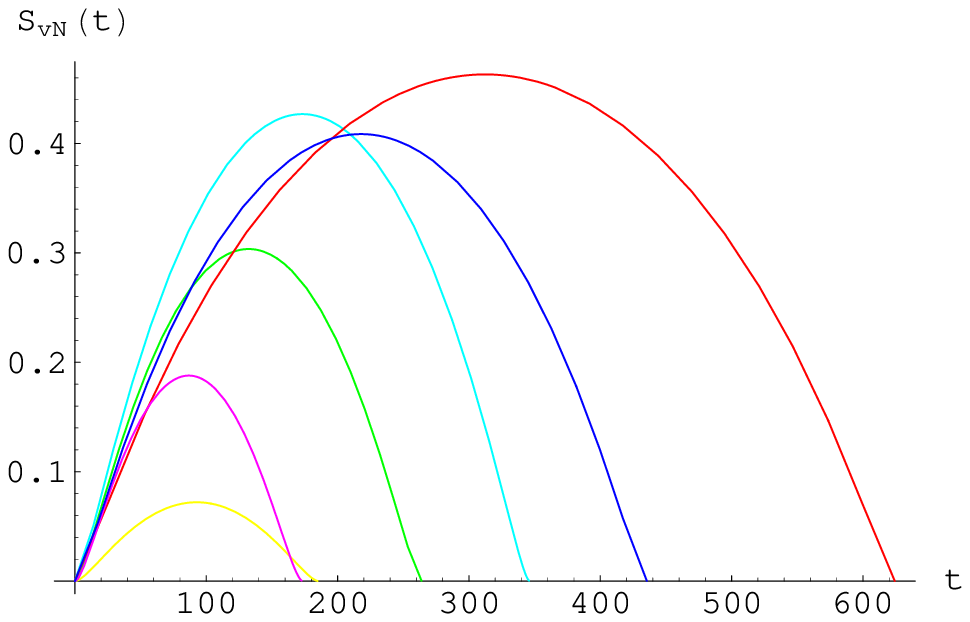}
\end{center}
\caption{von Neumann entropy versus optimized time (point (B-3)).}
\end{figure}
\par\begin{figure}[H]
\begin{center}
\includegraphics[width=8cm]{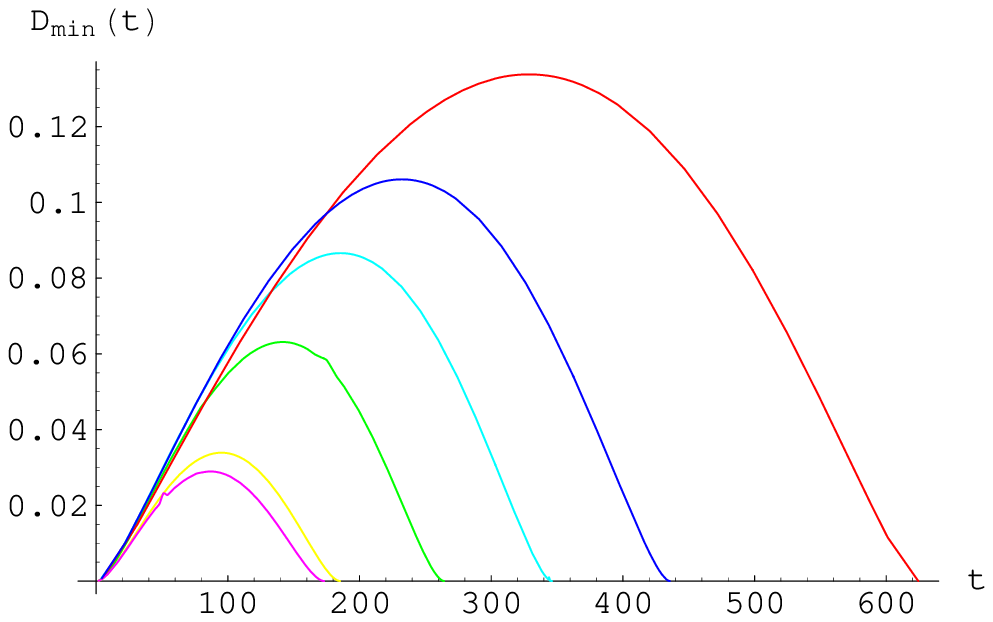}
\end{center}
\caption{Normalized geometric separation versus optimized time (point (B-3)).}
\end{figure}
\newpage

\subsection{Deutsch Algorithm for two qubits}

The running times are (point (A-1)):
\bea
T_{run}\Big|_{\rm unoptimized} &&:~~ \{T_{\rm magenta} = 183,~ T_{\rm cyan} = 128,~ T_{\rm blue} =97,~ T_{\rm green} = 67,~ T_{\rm yellow} = 55,~ T_{\rm red} = 52\}\nonumber\\
T_{run}\Big|_{\rm optimized} &&:~~ \{T_{\rm magenta} = 118,~ T_{\rm cyan} = 93,~ T_{\rm blue} =76,~ T_{\rm green} = 57,~ T_{\rm yellow} = 49,~ T_{\rm red} = 46\}\nonumber
\eea
\par\begin{figure}[H]
\begin{center}
\includegraphics[width=8cm]{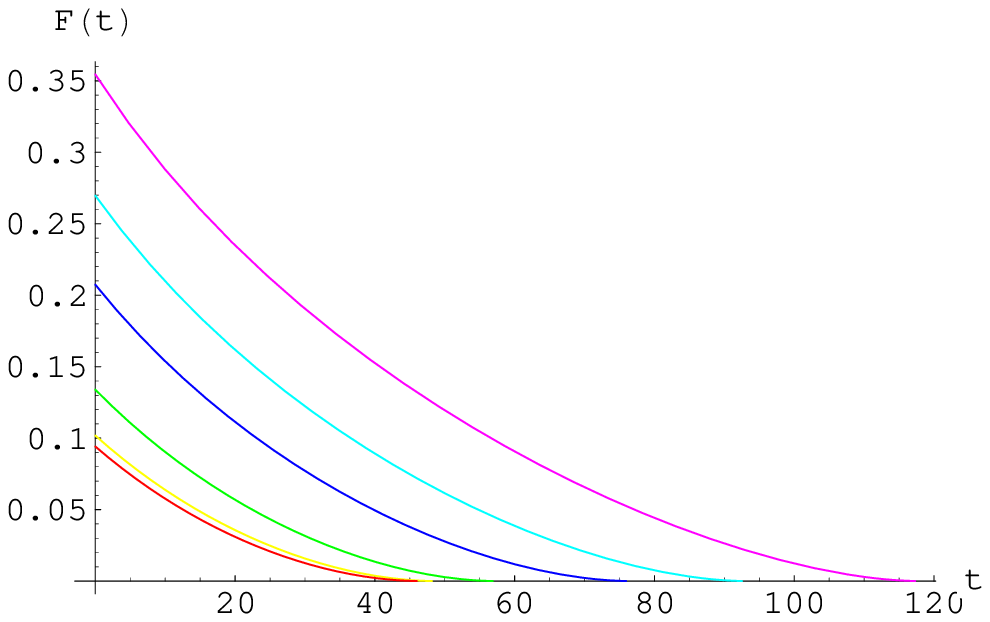}
\end{center}
\caption{Largest initial fidelity corresponds to the largest running time (point (A-2)).}
\end{figure}
\par\begin{figure}[H]
\begin{center}
\includegraphics[width=8cm]{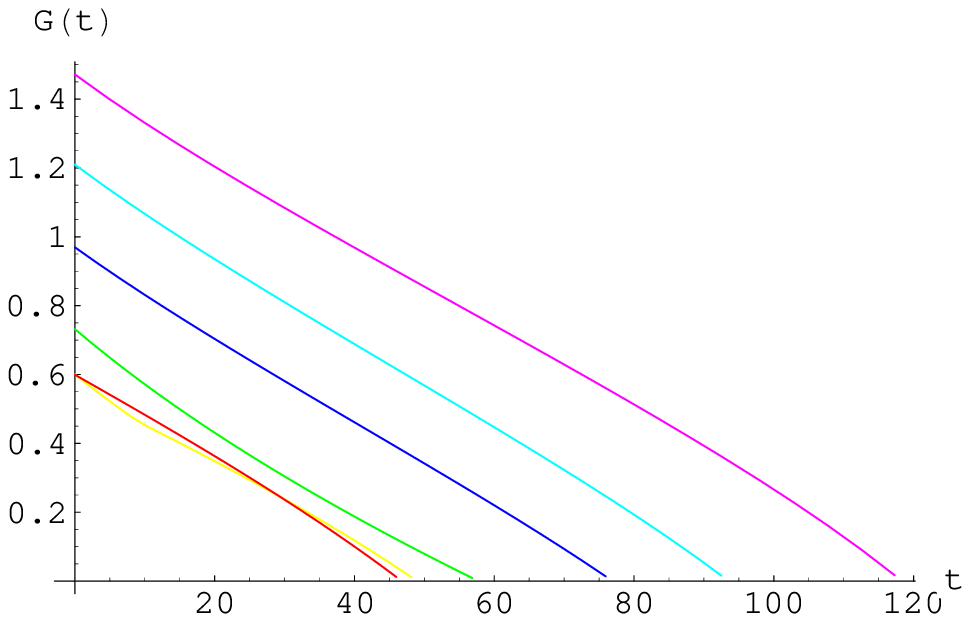}
\end{center}
\caption{Largest initial geometric separation corresponds to the largest running time (point (A-3)).}
\end{figure}
\par\begin{figure}[H]
\begin{center}
\includegraphics[width=8cm]{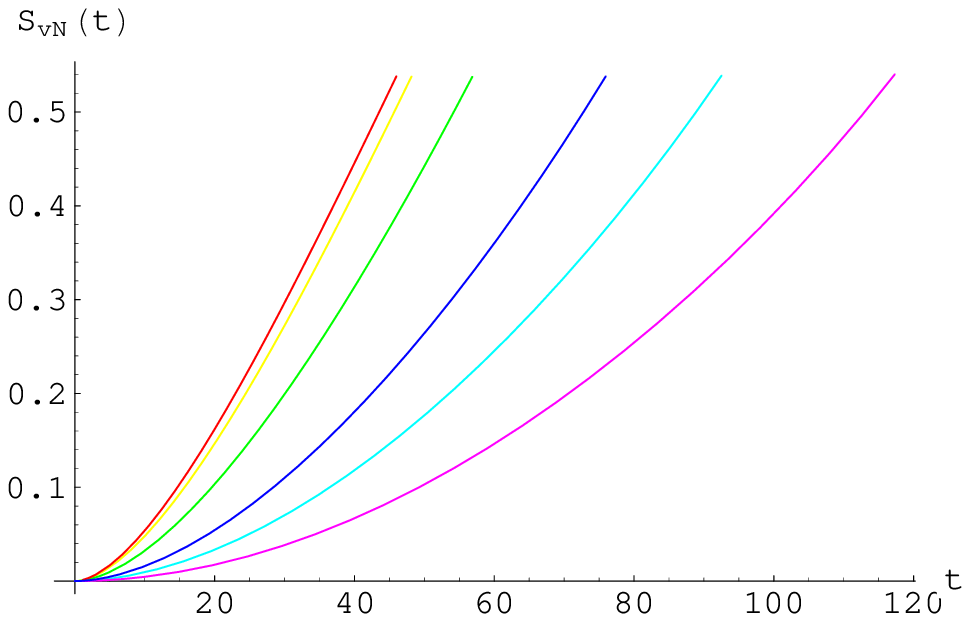}
\end{center}
\caption{States that generate entropy more quickly have a shorter running time (point (C-1)).}
\end{figure}
\par\begin{figure}[H]
\begin{center}
\includegraphics[width=8cm]{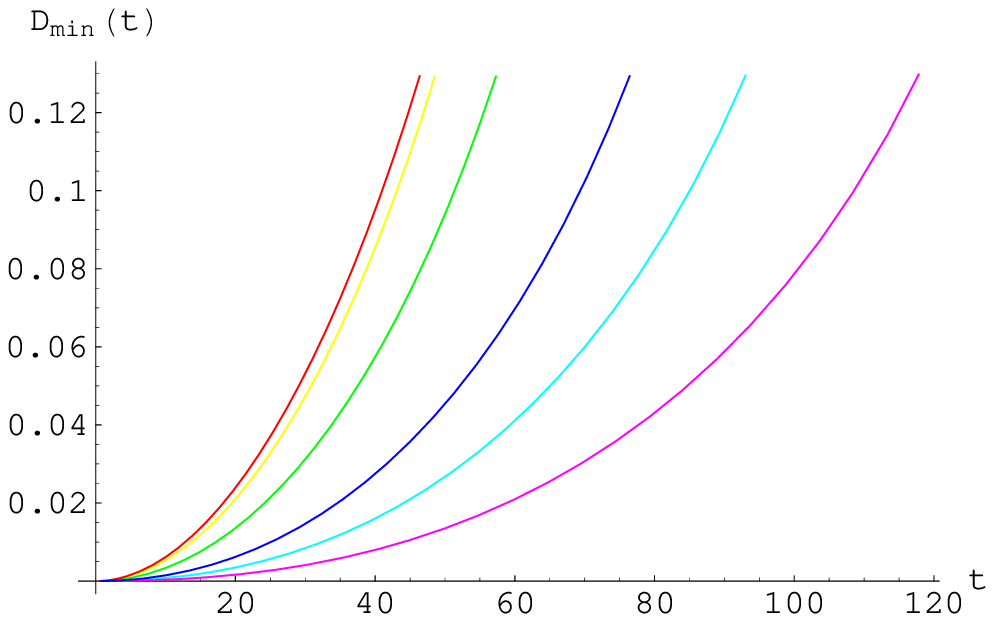}
\end{center}
\caption{States that generate normalized geometric entanglement more quickly have a shorter running time (point (C-1)).}
\end{figure}

\newpage

\subsection{Deutsch Algorithm for three qubits}

The running times are (point (A-1)):
\bea
T_{run}\Big|_{\rm unoptimized} &&:~~ \{T_{\rm magenta} = 217,~ T_{\rm yellow} = 104,~ T_{\rm red} =55,~ T_{\rm blue} = 43,~ T_{\rm green} = 40,~ T_{\rm cyan} = 37\}\nonumber\\
T_{run}\Big|_{\rm optimized} &&:~~ \{T_{\rm magenta} = 131,~ T_{\rm yellow} = 81,~ T_{\rm red} =49,~ T_{\rm blue} = 39,~ T_{\rm green} = 37,~ T_{\rm cyan} = 35\}\nonumber
\eea
\par\begin{figure}[H]
\begin{center}
\includegraphics[width=8cm]{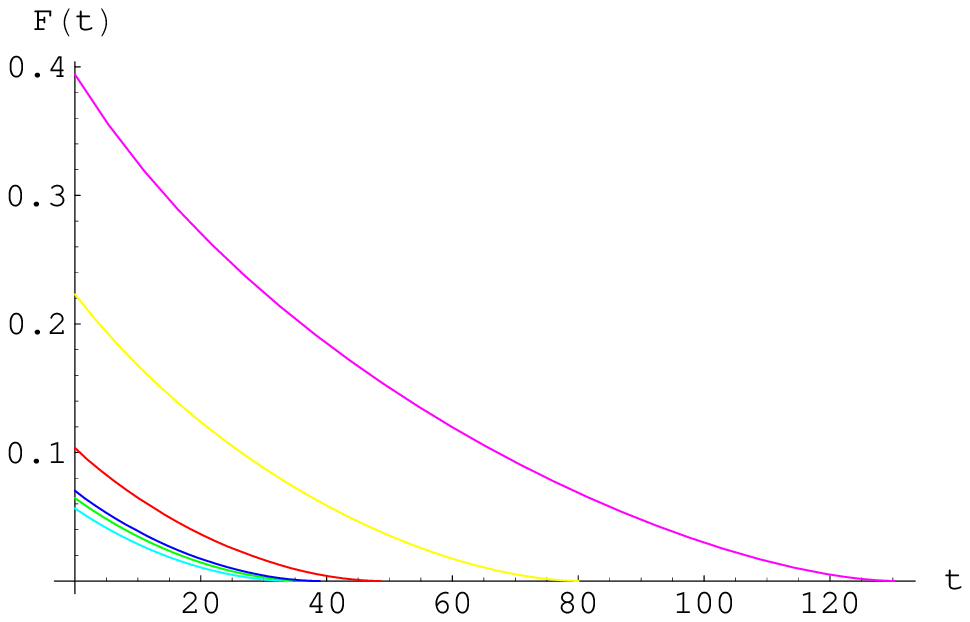}
\end{center}
\caption{Largest initial fidelity corresponds to the largest running time (point (A-2)).}
\end{figure}
\par\begin{figure}[H]
\begin{center}
\includegraphics[width=8cm]{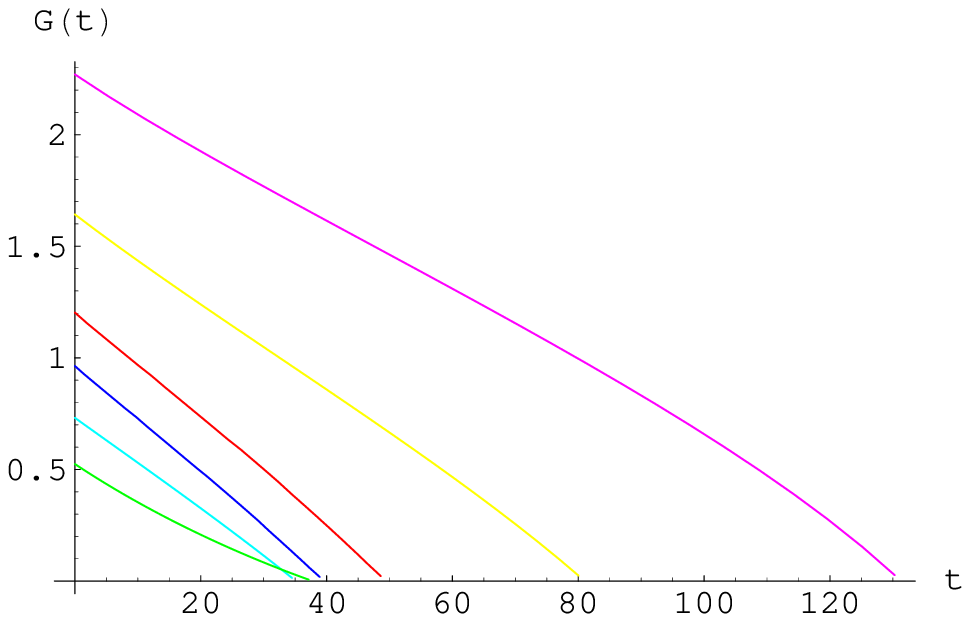}
\end{center}
\caption{Largest initial geometric separation corresponds to the largest running time (point (A-3)).}
\end{figure}
\par\begin{figure}[H]
\begin{center}
\includegraphics[width=8cm]{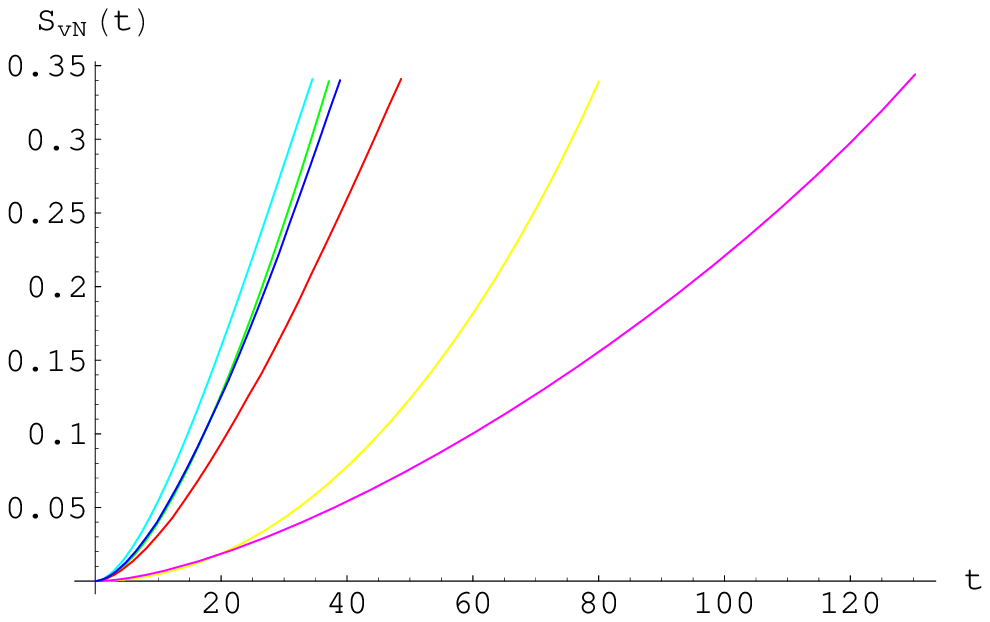}
\end{center}
\caption{States that generate entropy more quickly have a shorter running time (point (C-1)).}
\end{figure}
\par\begin{figure}[H]
\begin{center}
\includegraphics[width=8cm]{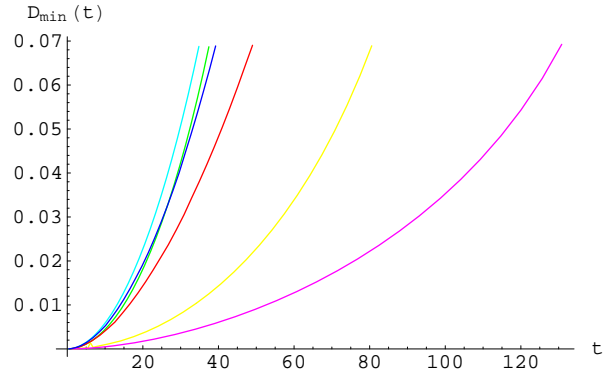}
\end{center}
\caption{States that generate geometric entanglement more quickly have a shorter running time (point (C-1)).}
\end{figure}

\newpage

\subsection{Constant-Time-Deutsch Algorithm for two qubits}

The running times are (point (A-1)):
\bea
T_{run}\Big|_{\rm unoptimized} &&:~~ \{T_{\rm cyan} = 165,~ T_{\rm blue} = 150,~ T_{\rm yellow} = 146\}\nonumber\\
T_{run}\Big|_{\rm optimized} &&:~~ \{T_{\rm cyan} = 110,~ T_{\rm blue} = 103,~ T_{\rm yellow} =102\}\nonumber
\eea
\par\begin{figure}[H]
\begin{center}
\includegraphics[width=8cm]{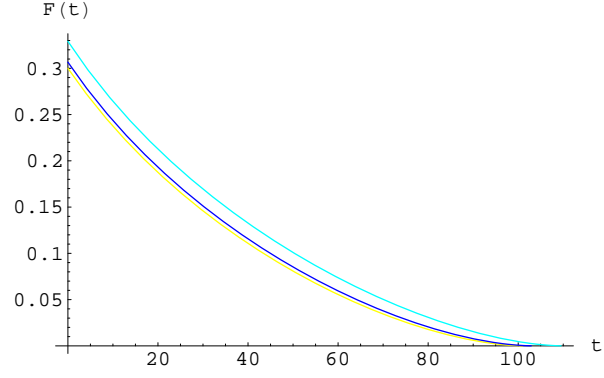}
\end{center}
\caption{Largest initial fidelity corresponds to the largest running time (point (A-2)).}
\end{figure}
\par\begin{figure}[H]
\begin{center}
\includegraphics[width=8cm]{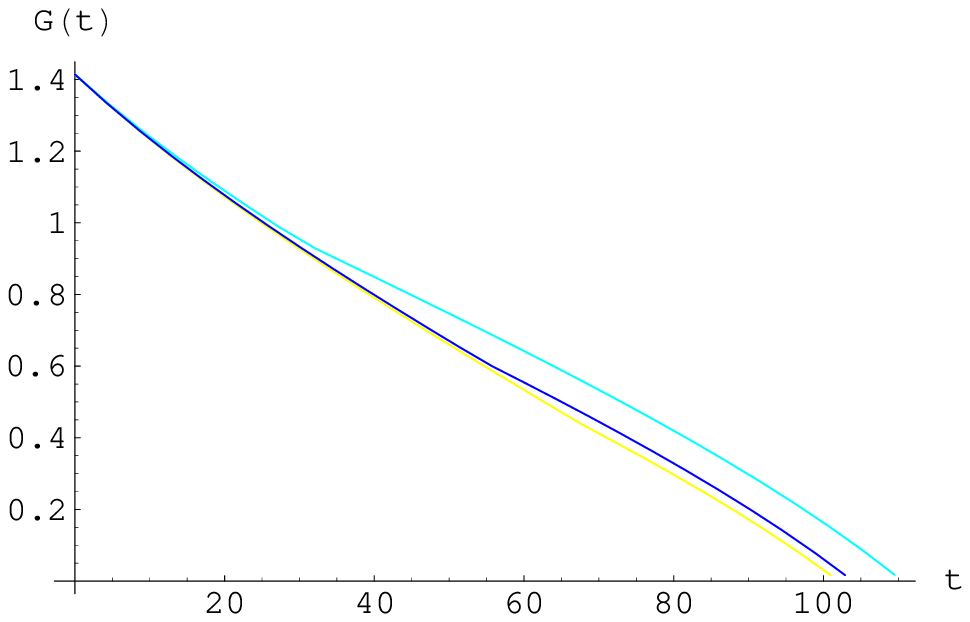}
\end{center}
\caption{Largest initial geometric separation corresponds to the largest running time (point (A-3)).}
\end{figure}
\par\begin{figure}[H]
\begin{center}
\includegraphics[width=8cm]{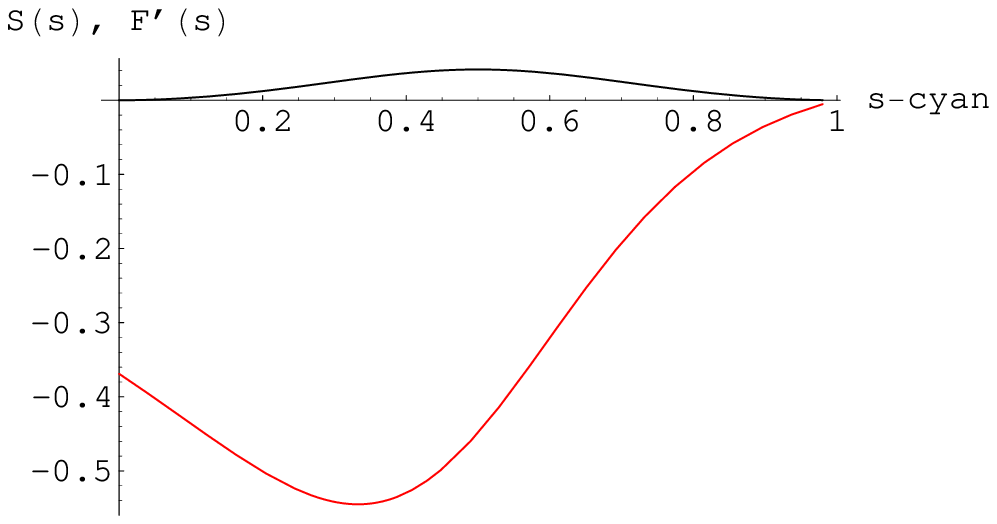}
\end{center}
\caption{The region of greatest magnitude of the rate of change of the fidelity  corresponds to the region of greatest entanglement (point (D-1)). This graph is for the cyan initial state.}
\end{figure}
\par\begin{figure}[H]
\begin{center}
\includegraphics[width=8cm]{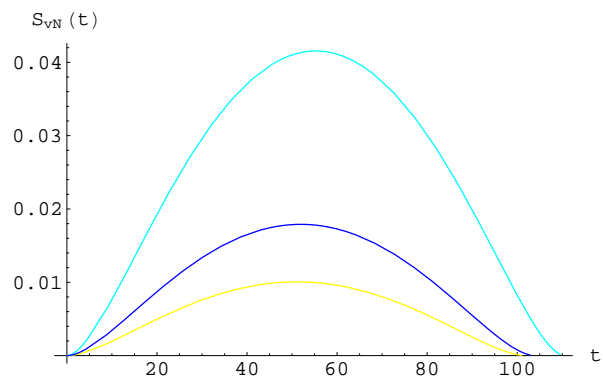}
\end{center}
\caption{von Neumann entropy versus optimized time (point (D-2)).}
\end{figure}

\section{Discussion and Outlook}
\label{conc}

The purpose of this paper is to study the role of entanglement in quantum computing. 
We have used the adiabatic formulation of quantum computing, so that we can explicitly track the time evolution of the quantities we are interested in. We have studied two different algorithms, adiabatic quantum search and the adiabatic Deutsch-Jozsa algorithm. We have also considered a variation of the Deutsch-Jozsa algorithm  \cite{Wei0512008} which we call the constant-time-Deutsch-Jozsa algorithm. We have introduced a new function to describe the physical property of entanglement and compared with the results for the entropy of entanglement. We also introduced a different function characterizing the distance between states and compared the results with the traditional fidelity. We have calculated the evolution of these quantities as a function of time, and also as a function of an optimized time variable, which corresponds to applying the adiabatic approximation in a maximally efficient way  \cite{rc0302138}.

For the search algorithm, we find that the region where the magnitude of the rate of change of the fidelity is greatest corresponds to the region of greatest entropy production. This result indicates that the state approaches  the final state most quickly when the entanglement is greatest, which supports the idea that entanglement is a resource for speed-up in quantum computation. 
We note that there also seems to be a correlation between greater maximum entanglement and longer running times. Although this result appears to contradict the conclusion above, we believe that it is an artifact of the structure of the adiabatic computation. 
We expect that both the initial fidelity, and the amount of entanglement that is produced during the evolution, will affect the running time of the algorithm.
However, since the Hamiltonian, and therefore the algorithm, depends explicitly on the choice of initial state, we are unable to study the roles of these factors independently. In other words, we believe that the longer running times produced in the cases we have studied are the result of larger initial fidelity, and not driven by larger maximum entanglement.

The results for the constant-time-Deutsch-Jozsa algorithm are similar to the results for the search algorithm.

The Deutsch-Jozsa algorithm is different from the previous two cases in that the entanglement of the final state is not zero. In this case, using the optimized time variable, we find that the states that generate entropy more quickly have a shorter running time. This result also seems to support the idea that entanglement is a resource for speed-up in quantum computation. 

In this paper we have worked only with $n=2$ and $n=3$. The next step is to extend our calculations to systems with larger $n$. Work on this is in progress.

\appendix
\section{Notation and Computational Basis}\label{AppendixA}
The basis states for one qubit in the $S_z$ representation are
\begin{eqnarray}
 |0\rangle& \doteq& \colzero\\
 |1\rangle&\doteq& \colone\nonumber ,
\end{eqnarray}
where we use the symbol $\doteq$ to denote a representation.
In the computational basis, an n-qubit state is written as
\begin{equation}
 |x \rangle_n=|x_0 x_1 ... x_n \rangle
\end{equation}
where $x_i =0 ~\mbox{or}~ 1 ~\forall~ i$, and the string they form is the binary representation of $x$.
We explicitly write the index $n$ here to denote the number of qubits. The dimension of the Hilbert space for a system of $n$ qubits is $N=2^n$.\\

The computational basis for $n=2$ qubits is:
\begin{eqnarray}
\label{comp2}
|0\rangle_2 = |00\rangle = |0\rangle_1 \otimes |0\rangle_1= (1,0,0,0)\\
|1\rangle_2 = |01\rangle = |0\rangle_1 \otimes |1\rangle_1 = (0,1,0,0)\nonumber\\
|2\rangle_2 = |10\rangle = |1\rangle_1 \otimes |0\rangle_1 = (0,0,1,0)\nonumber\\
|3\rangle_2 = |11\rangle = |1\rangle_1 \otimes |1\rangle_1 = (0,0,0,1)\nonumber
\end{eqnarray}
A general state of two qubits is written as
\begin{equation}
\label{2-gen}
|\Psi\rangle_2 = \sum_{i=0}^3 c_i |i\rangle_2 =  (c_0,~c_1,~c_2,~c_3)
\end{equation}
The computational basis for $n=3$ qubits is:
\bea
\label{comp3}
|0\rangle_3 = |000\rangle = |0\rangle_1 \otimes |0\rangle_1 \otimes |0\rangle_1 =(1,0,0,0,0,0,0,0)\\
|1\rangle_3 = |001\rangle = |0\rangle_1 \otimes |0\rangle_1 \otimes  |1\rangle_1 =(0,1,0,0,0,0,0,0)\nonumber\\
|2\rangle_3 = |010\rangle = |0\rangle_1 \otimes |1\rangle_1 \otimes |0\rangle_1  =(0,0,1,0,0,0,0,0)\nonumber\\
|3\rangle_3 = |011\rangle = |0\rangle_1 \otimes |1\rangle_1 \otimes |1\rangle_1 =(0,0,0,1,0,0,0,0)\nonumber\\
|4\rangle_3 = |100\rangle = |1\rangle_1 \otimes |0\rangle_1 \otimes |0\rangle_1 = (0,0,0,0,1,0,0,0)\nonumber\\
|5\rangle_3 = |101\rangle = |1\rangle_1 \otimes |0\rangle_1 \otimes |1\rangle_1 = (0,0,0,0,0,1,0,0)\nonumber\\
|6\rangle_3 = |110\rangle = |1\rangle_1 \otimes |1\rangle_1 \otimes |0\rangle_1 = (0,0,0,0,0,0,1,0)\nonumber\\
|7\rangle_3 = |111\rangle = |1\rangle_1 \otimes |1\rangle_1 \otimes |1\rangle_1 = (0,0,0,0,0,0,0,1)\nonumber
\eea
A general state of three qubits is written as
\begin{equation}
\label{3-gen}
|\Psi\rangle_3 = \sum_{i=0}^7 c_i |i\rangle_3=(c_0,~c_1,~c_2,~c_3,~c_4,~c_5,~c_6,~c_7)
\end{equation}

\section{Density matrices and reduced density matrices}\label{AppendixB}
The density matrix for an ensemble of pure states $ \{p_i,|\psi_i\rangle \}$ is
\begin{equation}
 \rho=\sum_{i} p_{i} |\psi_i\rangle  \langle \psi_i|
\end{equation}
meaning that the system is in state $|\psi_i\rangle$ with probability $p_i$. Recall that $Tr \rho=\sum_i p_i=1$ and $Tr \rho^2 \geq 1$. Since $\rho$ is positive, it must have a spectral decomposition
\begin{equation}
 \rho=\sum_j\lambda_j|j\rangle\langle j|,
\end{equation}
where the vectors $|j\rangle$ are orthogonal, and $\lambda_j$ are real, non-negative eigenvalues of $\rho$ (cf., e.g., \cite{Nielsen}).
 
For the special case of a pure state (or ensemble), one of the $p_{i}$ is unity and all others are zero. The density matrix reduces to
\bea
\label{rhopure}
\rho^{\rm pure}=|\psi_i\rangle \langle \psi_i|.
\eea
In this case only, $Tr(\rho^{\rm pure})^2 =1$. The eigenvalues of the pure state density matrix are $0$ and $1$.

The reduced density matrix is used for the description of subsystems of a composite quantum system. Assume we have a composite system with the subsystems $A$ and $B$. It is described by the density matrix $\rho^{AB}$. The reduced density operator for system $A$ is defined as
\begin{equation}
\rho^A=Tr_B(\rho^{AB}),
\end{equation}
where $Tr_B$ is the partial trace over systems $B$ (cf.,e.g.,\cite{Nielsen}).
\subsection{Calculations for $n=2$ qubits}

We construct the density matrix for a pure state of two qubits (i.e. a composite system) using (\ref{2-gen}) and (\ref{rhopure}). The density matrix is
\begin{equation}
\label{rho2}
 \rho_2=|\Psi\rangle_{2\;2}\langle\Psi| = \sum_{i,j}c_i c_j^*|i\rangle\langle j|\doteq\left (
 \begin{array}{cccc}c_0 c_0^*&c_0c_1^*&c_0c_2^*&c_0c_3^*\\
                                   c_1c_0^*&c_1c_1^*&c_1c_2^*&c_1c_3^*\\
                                   c_2c_0^*&c_2c_1^*&c_2c_2^*&c_2c_3^*\\
                                   c_3c_0^*&c_3c_1^*&c_3c_2^*&c_3c_3^*
  \end{array}\right)
  \end{equation}
We obtain the reduced density matrix for one of the subsystems by tracing out the degrees of freedom corresponding to the other qubit:
\begin{equation}
 \rho_2^{red}\doteq \left (
 \begin{array}{cc}c_0c_0^*+c_1c_1^*&c_0c_2^*+c_1c_3^*\\
                               c_2c_0^*+c_3c_1^*&c_2c_2^*+c_3c_3^*
  \end{array}\right )
\end{equation}
This reduced density matrix has the eigenvalues
\begin{equation}
\label{2-mu}
\mu_{\pm}=\frac{1}{2}(1\pm\sqrt{1-4|c_0 c_3 - c_1 c_2|^2})
\end{equation}
with $\mu_+ +\mu_- =1$. 
Note that for the case of a two qubit system, the reduced density matrices obtained by tracing out either of the two subsystems are equal. 
\subsection{Calculations for $n=3$ qubits}
For a pure composite state of three qubits we use (\ref{3-gen}) and (\ref{rhopure}) to get:
\begin{equation}
\label{rho3}
 \rho_{3}=|\Psi\rangle_{3\;3}\langle\Psi|
         =\sum_{i,j}c_i c_j^*|i\rangle_{3\:3}\langle j|\doteq\left (
             \begin{array}{cccccccc}c_0
	      c_0^*&c_0c_1^*&c_0c_2^*&c_0c_3^*&c_0c_4^*&c_0c_5^*&c_0c_6^*&c_0c_7^*\\
                                   c_1c_0^*&c_1c_1^*&c_1c_2^*&c_1c_3^*&c_1c_4^*&c_1
				   c_5^*&c_1c_6^*&c_1c_7^*\\
				   
c_2c_0^*&c_2c_1^*&c_2c_2^*&c_2c_3^*&  c_2c_4^*&c_2c_5^*&c_2c_6^*&c_2c_7^*\\
c_3c_0^*&c_3c_1^*&c_3c_2^*&c_3c_3^*& c_3c_4^*&c_3c_5^*&c_3c_6^*&c_3c_7^*\\
c_4c_0^*&c_4c_1^*&c_4c_2^*&c_4c_3^*& c_4c_4^*&c_4c_5^*&c_4c_6^*&c_4c_7^*\\
c_5c_0^*&c_5c_1^*&c_5c_2^*&c_5c_3^*& c_5c_4^*&c_5c_5^*&c_5c_6^*&c_5c_7^*\\
c_6c_0^*&c_6c_1^*&c_6c_2^*&c_6c_3^*& c_6c_4^*&c_6c_5^*&c_6c_6^*&c_6c_7^*\\
c_7c_0^*&c_7c_1^*&c_7c_2^*&c_7c_3^*& c_7c_4^*&c_7c_5^*&c_7c_6^*&c_7c_7^*
  \end{array}\right)
  \end{equation}
For the three qubit system we obtain the reduced density matrix by tracing out the degrees of freedom corresponding to the second and third qubits. We obtain:
\bea
 \rho_3^{red}\doteq\left(
\begin{array}{cc}
c_0 c_0^*+c_1 c_1^*+c_2 c_2^*+c_3 c_3^* & c_0 c_4^* +c_1 c_5^* +c_2 c_6^* +c_3 c_7^* \\
c_4 c_0^*+c_5 c_1^* +c_6 c_2^*+c_7 c_3^* & c_4 c_4^*+c_5 c_5^*+c_6 c_6^*+c_7 c_7^*
\end{array}
\right)
\eea
The eigenvalues are:
\bea
\label{3-mu}
&& \mu_\pm = \frac{1}{2}\left(1\pm\sqrt{1-4 XX}\right)\,;~~\mu_+ +\mu_- = 1\\
&& XX = (V_1 \cdot V_1^*)(V_2 \cdot V_2^*)-(V_1\cdot V_2^*)(V_2\cdot V_1^*)\nonumber\\
&& V_1 = (c_0,c_1,c_2,c_3)\nonumber\\
&& V_2 = (c_4,c_5,c_6,c_7)\nonumber
\eea
(Note that this formula also works for the two qubit case with the notation $V_1 = (c_0,c_1)$ and $V_2 = (c_2,c_3)$.)

If we trace out the first and second qubit, the reduced density matrix becomes
\bea
 \tilde{\rho}_3^{red}\doteq\left(
\begin{array}{cc}
c_0 c_0^*+c_2 c_2^*+c_4 c_4^*+c_6 c_6^* & c_0 c_1^* +c_2 c_3^* +c_4 c_5^* +c_6 c_7^* \\
c_1 c_0^*+c_3 c_2^* +c_5 c_4^*+c_7 c_6^* & c_1 c_1^*+c_3 c_3^*+c_5 c_5^*+c_7 c_7^*
\end{array}
\right)
\eea
In this case, the eigenvalues are given by Eq. (\ref{3-mu}) as well, but with
\bea
\tilde{V}_1 = (c_0,c_2,c_4,c_6)\\
\tilde{V}_2 = (c_1,c_3,c_5,c_7)\nonumber .
\eea

If we trace out the first and third qubit, the reduced density matrix becomes
\bea
 \tilde{\tilde{\rho}}_3^{red}\doteq\left(
\begin{array}{cc}
c_0 c_0^*+c_1 c_1^*+c_4 c_4^*+c_5 c_5^* & c_0 c_2^* +c_1 c_3^* +c_4 c_6^* +c_5 c_7^* \\
c_2 c_0^*+c_3 c_1^* +c_6 c_4^*+c_7 c_5^* & c_2 c_2^*+c_3 c_3^*+c_6 c_6^*+c_7 c_7^*
\end{array}
\right)
\eea
and the eigenvalues are given by Eq. (\ref{3-mu}) with
\bea
\tilde{\tilde{V}}_1 = (c_0,c_1,c_4,c_5)\\
\tilde{\tilde{V}}_2 = (c_2,c_3,c_6,c_7)\nonumber .
\eea
\section{Consistency condition for an unnormalized three qubit product state}\label{AppendixC}
The general expression for a three qubit product state is given by (compare (\ref{2qubit})):
\begin{eqnarray}
\label{3qubit}
 |p\rangle & = & (a_1|0\rangle +a_2|1\rangle)\otimes (b_1|0\rangle +b_2|1\rangle)\otimes (e_1|0\rangle
 +e_2|1\rangle)\label{3qubit_product}\\
 & = & \left\{a_1 b_1 e_1,a_1 b_1 e_2,a_1 b_2 e_1,a_1 b_2
   e_2,a_2 b_1 e_1,a_2 b_1 e_2,a_2 b_2 e_1,a_2 b_2
   e_2\right\}\nonumber
\end{eqnarray}
We do not assume the product state is normalized. The geometric distance between this state and an arbitrary
state ($c_0, c_1, c_2,c_3,c_4,c_5,c_6,c_7$) is (compare (\ref{distance})):
\begin{eqnarray}
D &=& (a_1b_1e_1 - c_0)^2 + (a_1b_1e_2 - c_1)^2 +
(a_1b_2e_1 - c_2)^2  + (a_1b_2e_2 - c_3)^2 \label{distance3qubit}\\
&+& (a_2b_1e_1 - c_4)^2 + (a_2b_1e_2 - c_5)^2 +
(a_2b_2e_1 - c_6)^2  + (a_2b_2e_2 - c_7)^2\nonumber
\end{eqnarray}
We can minimize this expression with respect to the parameters
($a_1, a_2, b_1, b_2, e_1, e_2$). Doing so results
in a set of 6 equations, for which one can find three
consistency equations by demanding that non--trivial
solutions exist. Defining
\begin{equation}
N_a = a_1^2 + a_2^2, \qquad N_b = b_1^2 + b_2^2, \qquad
N_e = e_1^2 + e_2^2,
\end{equation}
these equations are
\begin{eqnarray}
0 &=& 
N_e^4 N_a^2 N_b^2 \\
&-&
N_e^2 N_a N_b  [ (c_4^2+c_2^2+c_6^2+c_0^2) e_1^2 + ((c_1 c_0+c_7 c_6+c_3 c_2+c_5 c_4) e_1  e_2) +(c_1^2+c_5^2+c_3^2+c_7^2) e_2^2 ]\nonumber \\
&+&\left((c_4 c_2-c_0 c_6) e_1^2+(c_3 c_4+c_5 c_2-c_6 c_1-c_7 c_0) 
e_1 e_2+(c_5 c_3-c_1 c_7) e_2^2\right)^2
\nonumber \\
0 &=& 
N_a^4 N_b^2 N_e^2\nonumber\\
&-&
N_a^2 N_b N_e [(c_1^2+c_2^2+c_3^2+c_0^2) a_1^2 + ((c_4 c_0+c_6 c_2+c_5 c_1+c_3 c_7) a_1 a_2 )+(c_7^2+c_5^2+c_4^2+c_6^2)a_2^2]\nonumber\\
&+&
\left((-c_2 c_1+c_3 c_0) a_1^2+(-c_1 c_6+c_0 c_7+c_3 c_4-c_2 c_5) a_1 a_2
+(c_4 c_7-c_6 c_5) a_2^2\right)^2
\nonumber\\
0 &=&
N_b^4 N_a^2 N_e^2\nonumber\\
&-&
N_b^2 N_a N_e[(c_4^2+c_5^2+c_1^2+c_0^2)b_1^2+ ((c_7 c_5+c_2 c_0+c_3 c_1+c_4 c_6) b_1 b_2)+  b_1^2+ (c_2^2+c_6^2+c_7^2+c_3^2) b_2^2  ]\nonumber\\
&+&
\left((-c_4 c_1+c_5 c_0) b_1^2+(-c_6 c_1+c_0 c_7-c_4 c_3+c_2 c_5) b_1 b_2+
(c_2 c_7-c_3 c_6) b_2^2\right)^2\nonumber
\end{eqnarray}
\par\noindent
The relationship of these consistency conditions to the different ways of tracing
out states when defining the von Neumann entropy is currently under investigation.

\section{Analytic Solutions for Minimum Distance to a normalized two qubit Product State}
\label{AppendixD}

For $n=2$ the general product state is given in (\ref{2qubit}).
 If we use the change of variables:
\begin{eqnarray}
a_{1}=\sqrt{N_{a}}\cos\theta &  & a_{2}=\sqrt{N_{a}}\sin\theta\label{change_of_variables}\\
b_{1}=\sqrt{N_{b}}\cos\phi &  & b_{2}=\sqrt{N_{b}}\sin\phi\nonumber 
\end{eqnarray}
 where:
\begin{eqnarray*}
N_{a}=a_{1}^{2}+a_{2}^{2} & \mbox{ and } & N_{b}=b_{1}^{2}+b_{2}^{2}.
\end{eqnarray*}
 the product state becomes: 
\begin{equation}
\sqrt{N_{a}N_{b}}\left( \cos\theta\cos\phi,\,\cos\theta\sin\phi,\,\sin\theta\cos\phi,\,\sin\theta\sin\phi\right) .
\end{equation}
 The direction of the state 4-vector is fully determined by $\theta$
and $\phi$. The normalizations $N_{a}$ and $N_{b}$ of the individual
states determine the length of the state 4-vector. The
normalization condition for the product state is $N_{a}N_{b}=1$. 

The distance between a general normalized state:
\begin{equation}
\left\{ c_0,\, c_1,\, c_2,\, c_3\right\} \qquad\mbox{ where }\qquad c_3^{2}=1-c_0^{2}-c_1^{2}-c_2^{2}
\end{equation}
 and a general product state is:
\begin{equation}
\langle d|d\rangle = \left(c_0-a_{1}b_{1}\right)^{2}+\left(c_1-a_{1}b_{2}\right)^{2}+\left(c_2-a_{2}b_{1}\right)^{2}+\left(c_3-a_{2}b_{2}\right)^{2}.
\end{equation}
 Using the change of variables above leads to:
\begin{eqnarray}
\langle d|d\rangle &=&\left(c_0-\sqrt{N_{a}N_{b}}\cos\theta\cos\phi\right)^{2}+\left(c_1-\sqrt{N_{a}N_{b}}\cos\theta\sin\phi\right)^{2}\\
&+&\left(c_2-\sqrt{N_{a}N_{b}}\sin\theta\cos\phi\right)^{2}+\left(c_3-\sqrt{N_{a}N_{b}}\sin\theta\sin\phi\right)^{2}.\label{distance_in_angles}\nonumber
\end{eqnarray}
 In order to find the minimum distance to a product state we want
to find values for $\theta$ and $\phi$ such that the derivative
of this distance with respect to both variables ($\theta$ and $\phi$) vanishes. Differentiating
this equation with respect to $\theta$ and setting the derivative
equal to zero leads to the equation:
\begin{equation}
\tan\theta=\frac{c_2+c_3\tan\phi}{c_0+c_1\tan\phi}.\label{tan_theta}
\end{equation}
Differentiating this equation with respect to $\phi$ and setting
the derivative equal to zero leads to the equation:
\begin{equation}
\tan\phi=\frac{c_1+c_3\tan\theta}{c_0+c_2\tan\theta}.\label{tan_phi_1st}
\end{equation}
 Neither of these equations depends on the normalization $N_{a}N_{b}$
of the general product state. Solution of these equations therefore
leads to a line of product states with arbitrary length and the intersection of this line 
with the unit 4-sphere is the physical product state. Note also that when the state under 
consideration is a product state, it will satisfy $c_0c_3=c_1c_2$, and in this case
(\ref{tan_theta}) and (\ref{tan_phi_1st}) reduce to: 
\begin{equation}
\tan\theta=\frac{c_2}{c_0}\,;~~\tan\phi=\frac{c_1}{c_0}
\end{equation}
which is consistent with $c_0=a_{1}b_{1}$, $c_1=a_{1}b_{2}$ and $c_2=a_{2}b_{1}$, and the definitions in (\ref{change_of_variables}).

Substituting  (\ref{tan_theta}) into (\ref{tan_phi_1st})
and solving for $\tan\phi$ leads to the solutions:
\begin{equation}
\tan\phi=\frac{c_1^{2}+c_3^{2}-(c_0^{2}+c_2^{2})\pm\sqrt{4\left(c_0c_1+c_2c_3\right)^{2}+\left(c_1^{2}+c_3^{2}-(c_0^{2}+c_2^{2})\right)^{2}}}{2\left(c_0c_1+c_2c_3\right)}.\label{tan_phi}
\end{equation}
 For an arbitrary normalized state determined by $c_0,\, c_1$
and $c_2$, the closest product state can be determined
by (\ref{tan_theta}) and (\ref{tan_phi}). The distance to
this product state can then be determined from (\ref{distance_in_angles})
after choosing its normalization, $N_{a}N_{b}$.

\section{Equations for Minimum Distance to a normalized three qubit Product State}
\label{AppendixE}

For $n=3$ the general product state is given in (\ref{3qubit_product}).
 If we use the change of variables:
\begin{eqnarray}
a_{1}=\sqrt{N_{a}}\cos\theta &  & a_{2}=\sqrt{N_{a}}\sin\theta\label{change_of_variables_3qubit}\\
b_{1}=\sqrt{N_{b}}\cos\phi &  & b_{2}=\sqrt{N_{b}}\sin\phi\nonumber \\
e_{1}=\sqrt{N_{e}}\cos\psi &  & e_{2}=\sqrt{N_{e}}\sin\psi\nonumber 
\end{eqnarray}
 where:
\begin{eqnarray*}
N_{a}=a_{1}^{2}+a_{2}^{2} & ~~~~~ N_{b}=b_{1}^{2}+b_{2}^{2}~~~~~ & N_{e}=e_{1}^{2}+e_{2}^{2}.
\end{eqnarray*}
 the product state becomes: 
\begin{eqnarray}
\sqrt{N_{a}N_{b}N_{e}}\left( \cos\theta\cos\phi\cos\psi,\,\cos\theta\cos\phi\sin\psi,
                            \,\cos\theta\sin\phi\cos\psi,\,\cos\theta\sin\phi\sin\psi,\,\right.&& \\
\left. \sin\theta\cos\phi\cos\psi,\,\sin\theta\cos\phi\sin\psi,
        \, \sin\theta\sin\phi\cos\psi \, \sin\theta\sin\phi\sin\psi\right) . \nonumber
\end{eqnarray}
The direction of the state 8-vector is fully determined by $\theta$, $\phi$ and $\psi$. The normalizations $N_{a}$, $N_{b}$ and $N_{e}$ of the individual
states determine the length of the state 8-vector. The
normalization condition for the product state is $N_{a}N_{b}N_{e}=1$. 

The distance between a general normalized state and a product state is given in (\ref{distance3qubit}).
Using the change of variables above leads to:
\begin{eqnarray}
\langle d|d\rangle
&=&\left(c_0-\sqrt{N_{a}N_{b}N_{e}}\cos\theta\cos\phi\cos\psi\right)^{2}+\left(c_1-\sqrt{N_{a}N_{b}N_{e}}\cos\theta\cos\phi\sin\psi\right)^{2}\label{distance_in_angles_3qubit}\\
&+&\left(c_2-\sqrt{N_{a}N_{b}N_{e}}\cos\theta\sin\phi\cos\psi\right)^{2}+\left(c_3-\sqrt{N_{a}N_{b}N_{e}}\cos\theta\sin\phi\sin\psi\right)^{2}.\nonumber\\
&+&\left(c_4-\sqrt{N_{a}N_{b}N_{e}}\sin\theta\cos\phi\cos\psi\right)^{2}+\left(c_5-\sqrt{N_{a}N_{b}N_{e}}\sin\theta\cos\phi\sin\psi\right)^{2}.\nonumber\\
&+&\left(c_6-\sqrt{N_{a}N_{b}N_{e}}\sin\theta\sin\phi\cos\psi\right)^{2}+\left(c_7-\sqrt{N_{a}N_{b}N_{e}}\sin\theta\sin\phi\sin\psi\right)^{2}.\nonumber
\end{eqnarray}
In order to find the minimum distance to a product state we want
to find values for $\theta$, $\phi$ and $\psi$ such that the derivative
of this distance with respect to these variables ($\theta$, $\phi$ and $\psi$) vanishes. Differentiating
with respect to each variable and setting the derivative
equal to zero leads to the equations:
\begin{equation}
\tan\theta=\frac{c_4+c_6\tan\phi+c_5\tan\psi+c_7\tan\phi\tan\psi}
                {c_0+c_2\tan\phi+c_1\tan\psi+c_3\tan\phi\tan\psi}.\label{tan_theta_3qbit}
\end{equation}
\begin{equation}
\tan\phi=\frac{c_2+c_6\tan\theta+c_3\tan\psi+c_7\tan\theta\tan\psi}
                {c_0+c_4\tan\theta+c_1\tan\psi+c_5\tan\theta\tan\psi}.\label{tan_phi_3qbit}
\end{equation}
\begin{equation}
\tan\psi=\frac{c_1+c_5\tan\theta+c_3\tan\phi+c_7\tan\theta\tan\phi}
                {c_0+c_4\tan\theta+c_2\tan\phi+c_6\tan\theta\tan\phi}.\label{tan_psi_3qbit}
\end{equation}
None of these equations depends on the normalization $N_{a}N_{b}N_{e}$
of the general product state. Solution of these equations therefore
leads to a line of product states with arbitrary length and the intersection of this line with the 
unit 8-sphere is the physical product state. Solutions of these equations can be substituted
into (\ref{distance_in_angles_3qubit}) to find the distance to the nearest product state.

This set of equations is easily generalizable to the case for arbitrary $n$.
It is not clear that analytic solutions of these equations would be possible
for arbitrary $n$, but they can always be solved numerically. They can
be used to provide a measure of the entanglement which is symmmetric among the qubits.
The advantage of this approach is that it provides an unambigous measure of entanglement.
  

\end{document}